\documentclass[aip,amsmath,amssymb,reprint, twocolumn]{revtex4-1}
\usepackage[utf8]{inputenc}
\usepackage[T1]{fontenc}
\usepackage{mathptmx}
\usepackage{rotating}
\usepackage{longtable}
\usepackage{listings}
\usepackage{xcolor}
\usepackage{siunitx}
\usepackage{lipsum}
\usepackage{wrapfig}
\definecolor{codegreen}{rgb}{0,0.6,0}
\definecolor{codegray}{rgb}{0.5,0.5,0.5}
\definecolor{codepurple}{rgb}{0.58,0,0.82}
\definecolor{backcolour}{rgb}{0.95,0.95,0.92}
\lstdefinestyle{mystyle}{
    backgroundcolor=\color{backcolour},   
    commentstyle=\color{codegreen},
    keywordstyle=\color{magenta},
    numberstyle=\tiny\color{codegray},
    stringstyle=\color{codepurple},
    basicstyle=\ttfamily\footnotesize,
    breakatwhitespace=false,         
    breaklines=true,                 
    captionpos=b,                    
    keepspaces=true,                 
    numbersep=5pt,                  
    showspaces=false,                
    showstringspaces=false,
    showtabs=false,                  
    tabsize=2
}

\usepackage{times,amsmath}
\usepackage{epsfig}
\usepackage{color}
\usepackage{longtable}
\usepackage{ulem}
\usepackage{graphicx}
\usepackage{dcolumn}
\usepackage{bm}
\usepackage{bookmark}
\usepackage{tabularx}
\usepackage{hyperref}
\usepackage{multirow}
\usepackage{booktabs}
\hypersetup{colorlinks=true, citecolor=blue, filecolor=blue, linkcolor=blue, urlcolor=blue}

\usepackage[titles]{tocloft}
\begin{document}
\title{AI-Assisted Rapid Crystal Structure Generation Towards a Target Local Environment}

\author{Osman Goni Ridwan}
\affiliation{Department of Mechanical Engineering and Engineering Science, University of North Carolina at Charlotte, Charlotte, NC 28223, USA}

\author{Sylvain Pitié}
\affiliation{Applied Quantum Chemistry Group, Poitiers University-CNRS, Poitiers 86073, France}

\author{Monish Soundar Raj}
\affiliation{Department of Computer Science, University of North Carolina at Charlotte, Charlotte, NC 28223, USA}

\author{Dong Dai}
\affiliation{Department of Computer and Information Sciences, University of Delaware, Newark, DE 19716, USA}

\author{Gilles Frapper}
\email{gilles.frapper@univ-poitiers.fr}
\affiliation{Applied Quantum Chemistry Group, Poitiers University-CNRS, Poitiers 86073, France}

\author{Hongfei Xue}
\email{hongfei.xue@charlotte.edu}
\affiliation{Department of Computer Science, University of North Carolina at Charlotte, Charlotte, NC 28223, USA}

\author{Qiang Zhu}
\email{qzhu8@charlotte.edu}
\affiliation{Department of Mechanical Engineering and Engineering Science, University of North Carolina at Charlotte, Charlotte, NC 28223, USA}
\affiliation{North Carolina Battery Complexity, Autonomous Vehicle and Electrification (BATT CAVE) Research Center, Charlotte, NC 28223, USA}

\date{\today}
\begin{abstract}
In the field of material design, traditional crystal structure prediction approaches require structural sampling through computationally expensive energy minimization methods using either force fields or quantum mechanical simulations. While emerging artificial intelligence (AI) generative models have shown great promise in generating realistic crystal structures more rapidly, most existing models fail to account for the unique symmetries and periodicity of crystalline materials, and they are limited to handling structures with only a few tens of atoms per unit cell. Here, we present a symmetry-informed AI generative approach called Local Environment Geometry-Oriented Crystal Generator (\texttt{LEGO-xtal}) that overcomes these limitations. Our method generates initial structures using AI models trained on an augmented small dataset, and then optimizes them using machine learning structure descriptors rather than traditional energy-based optimization. We demonstrate the effectiveness of \texttt{LEGO-xtal} by expanding from 25 known low-energy sp$^2$ carbon allotropes to over 1,700, all within 0.5 eV/atom of the ground-state energy of graphite. This framework offers a generalizable strategy for the targeted design of materials with modular building blocks, such as metal–organic frameworks and next-generation battery materials.\\

\end{abstract}

\maketitle
\makeatletter
\tableofcontents

\section{INTRODUCTION}
For new materials design, the ability to predict the crystal structure of a material is crucial for understanding its properties and potential applications. However, crystal structure prediction (CSP) has long been considered a grand challenge in materials science. Analogous to protein structure prediction \cite{jumper2021highly}, solving a typical CSP problem requires searching for the minimum energy arrangement of 10–100 atoms within an unknown periodic crystal lattice \cite{Oganov-NRM-2019}. This formulation results in a combinatorial search space of astronomical size \cite{Pickard-JPCM-2011, Oganov-JCP-2006, Oganov-ACR-2011, Stillinger-PRE-1999, Martiniani-PRE-2016}. Additionally, evaluating each trial solution involves time-intensive quantum mechanical simulations, making the problem even more formidable.

Around the 2000s, the CSP field experienced a rapid development of non-empirical structure search algorithms by borrowing ideas from the global optimization community \cite{Oganov-NRM-2019}. Several algorithms—including random search \cite{Pickard-JPCM-2011}, genetic/evolutionary algorithms \cite{Oganov-JCP-2006, Wang-PRB-2010, Lonie-CPC-2011, GASP-Python}, simulated annealing \cite{Pannetier-Nature-1990}, and basin hopping \cite{banerjee2021crystal} were successfully applied to various material systems under both ambient and extreme conditions. These successes were largely attributed to strategies aimed at reducing dimensionality at the algorithmic level \cite{Pickard-JPCM-2011, Oganov-NRM-2019, han2025efficient}. While the number of possible structural arrangements is infinite, many can be immediately discarded due to unfavorable energy costs. By incorporating reasonable initial guesses (e.g., imposing chemical and symmetry constraints \cite{Pickard-JPCM-2011, han2025efficient}) and efficient structure-switching mechanisms (e.g., following physical vibration modes \cite{lyakhov2010predict, Zhu-CE-2012, QZhu-PRB-2015}, and self-learning intelligence \cite{Oganov-JCP-2006, Wang-PRB-2010}), researchers were able to focus on exploring low-energy basins with a much reduced computational cost. However, these traditional approaches remain heavily reliant on high-end electronic structure prediction methods, which come with a computational cost scaling relation $O(N^3)$ with respect to the number of valence electrons of the given system. As a result, a practical CSP task is limited to 20–30 atoms per periodic system, requiring several days to weeks on a single CPU node consisting of 48–96 cores.

Recently, the advent of artificial intelligence (AI) generative models has opened up new avenues for rapid crystal structure generation. Inspired by the successes of synthetic image generation and protein structure prediction \cite{jumper2021highly}, various AI-based generative models have been introduced to the materials community since 2018, including Generative Adversarial Networks (GANs) \cite{kim2020generative, zhu2024wycryst}, Variational Autoencoders (VAEs) \cite{noh2019inverse}, and more recent active learning \cite{merchant2023scaling}, 
diffusion \cite{CDVAE, DiffCSP, zeni2025generative, levy2024symmcd, diffcsp++, miller2024flowmm, zhong2025practical} and transformer models \cite{antunes2024crystal, gruver2024fine, cao2024space, kazeev2025wyckoff}. These models offer a fundamentally new approach to exploring structure space by learning data-driven representations, allowing for the efficient generation of low-energy crystal structures at a much faster rate than traditional global optimization methods. However, the direct application of AI-based models to the CSP task faces a significant challenge: the generated structures often lack diversity or fail to satisfy essential physical constraints, as these models tend to either replicate structural prototypes from the training data, or generate invalid and unstable structure, rather than producing truly novel and physical-plausible structures \cite{cheetham2024artificial}. This limitation arises because these data-driven models primarily generate crystal structures only based on the statistical patterns from the training data source without incorporating the unique characteristics of crystalline materials, particularly their symmetries. To address the challenge, several recent works have attempted to incorporate symmetry into generative AI model design with the aim of enhancing the quality and validity of generated structures \cite{diffcsp++, levy2024symmcd, antunes2024crystal, cao2024space, kazeev2025wyckoff, zhu2024wycryst}. However, most of the previous works simply add each individual symmetry representation into the existing AI models. Consequently, the learned models may be able to generate new candidate structures with a symmetry identical to that in the training samples. Nevertheless, they are less likely to generate those closely symmetry-related structures, thus largely limiting the model's learning efficiency. 

In this work, we present a novel symmetry-informed generative model called \texttt{LEGO-xtal} (Local Environment Geometry-Oriented Crystal Generator) that overcomes the limitations of existing AI models in crystal structure prediction. Unlike the previous AI generative models that focus on generation of relatively small crystal systems from a large dataset without the explicit constraints on the local chemical
environment, our approach is designed to generate complex crystal structures with well-defined local environments (i.e. specified chemical building blocks) while preserving the inherent symmetries and periodicity characteristic of crystalline materials. As a proof of concept, we train our generative models (VAEs and GANs) on an augmented dataset of known three-coordinated carbon allotropes —commonly referred to as sp$^2$ carbon structures— which enables the exploration of a broad range of structural configurations. Starting from only 25 known low-energy sp$^2$ carbon allotropes (as well as additional 115 high-energy examples), we demonstrate the effectiveness of our approach by generating over 1,700 new structures with varied dimensionality, topology and unit cell sizes. 

The remainder of this paper is organized as follows: First, we discuss the practical challenges in using AI models for crystal structure prediction and introduce our key solutions. Then, we present the detailed methodology of the \texttt{LEGO-xtal} framework. Finally, we analyze the generated structures and highlight several interesting findings.

\section{Computational Methodology}
Unlike computer vision objects like images and point clouds that only require limited local information, crystal structures exhibit distinct long-range translational and local rotational symmetries. These symmetry elements are rigorously defined through crystallographic space groups using group theory \cite{ITA}. When combined with specific chemical systems, crystal chemistry provides valuable rules and heuristics to guide our understanding of crystal packing and structure formation \cite{Oganov-NRM-2019}. To design an effective structure prediction model, we need to carefully consider how to incorporate these crystallographic and chemical principles into the modeling framework.

\subsection{Symmetry-based Crystal Representation}
\begin{figure*}[htbp]
    \centering
    \includegraphics[width=0.8\textwidth]{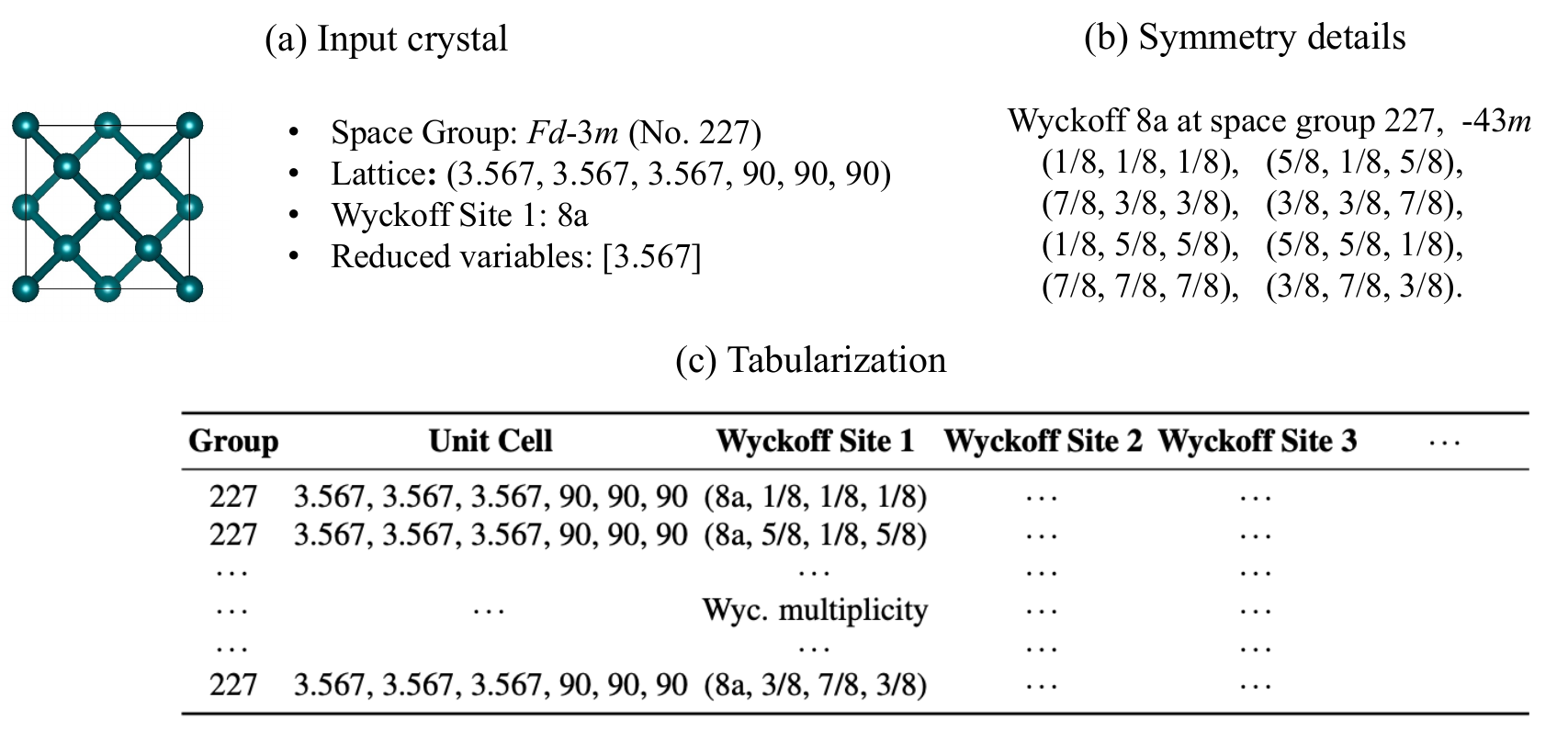}
    \caption{\textbf{Illustration of symmetry encoded crystal representation}. (a) shows the example diamond crystal structure, (b) lists the associated Wyckoff symmetry operations to describe the carbon atoms in the diamond; (c) lists the complete tabular representation of a diamond crystal according to its highest space group symmetry 227.}
    \label{fig:rep}
\vspace{-3mm}
\end{figure*}

To store and exchange crystal structure information, the Crystallographic Information File (CIF) is the most common and standardized text-based file format, containing atomic positions, unit cell parameters, and symmetry information \cite{CIF}. For example, the well-known diamond structure can be compiled into a tabular format as shown in Figure \ref{fig:rep}. Diamond crystallizes in space group \textit{Fd}$\bar{3}$\textit{m} ( SG 227), with a cubic unit cell ( a = b = c = 3.567 \AA, and $\alpha$ = $\beta$ = $\gamma$ = 90$^\circ$). In each unit cell, there are 8 symmetry-equivalent carbon atoms occupying an orbit, forming $\bar{4}3m$ site symmetry along the crystallographic [100], [111], and [110] directions \cite{ITA}. In crystallography, the Wyckoff site notation is used to describe the arrangement of atoms in a crystal structure. Each Wyckoff site is defined by its symmetry properties and multiplicity, which indicates how many equivalent positions exist for that site within the unit cell. The notation typically consists of a number indicating the multiplicity and a letter indicating the symmetry type (e.g., 8\textit{a}, 4\textit{b}, etc.). In this case, the diamond structure has 8 carbon atoms occupying Wyckoff site 8\textit{a}, which is characterized by its high symmetry (see Figure \ref{fig:rep}b).

Using this notation, it is straightforward to consider each structure in a tabular format with the following attributes.
\begin{itemize}
    \item \textit{Space Group number} which ranges from 1 to 230. 
    \item \textit{Cell Parameters}: $a, b, c, \alpha, \beta, \gamma$, which is a 6-vector that defines the cell lengths and angles in a standard crystallographic setting. 
    \item \textit{Wyckoff site 1}: (index, element type, $x$, $y$, $z$), which represents the index of Wyckoff site and the associated fractional coordinates bounded in the (0, 1) interval.
    \item \textit{Wyckoff site 2}: (index, element type, $x$, $y$, $z$). 
    \item $\cdots$
\end{itemize}

Thus, one essentially needs two sets of variables to describe a structure: (i) the \textit{discrete variables} denoting the exact space group symmetry and Wyckoff site indices; and (ii) the \textit{continuous variables} describing the cell parameters and Wyckoff positions. For high symmetry cases, the cell parameters and $(x,y,z)$ coordinates may have additional constraints. For instance, cubic unit cells must have $a=b=c$ and $\alpha=\beta=\gamma=90^\circ$. The Wyckoff position may be completely or partially restricted to specific fractional coordinates due to site symmetry requirements. If input variables do not satisfy symmetry requirements, deterministic algorithms can check if variables can be symmetrized within an acceptable threshold. Considering the symmetry constraints, the free continuous variables are thus reduced to a 1-length vector [3.567] for the diamond carbon.

Therefore, we can use a fixed-length 1D vector to represent a crystal structure with a maximum number of Wyckoff positions. For a maximum of 8 Wyckoff positions, the vector length is 1 + 6 + 8$\times$5 = 47. For a single-component system, this reduces to 39. 

Although this representation is straightforward, it is not mathematically unique. In principle, one crystal structure can be represented by many different 1D vectors by choosing different choices of $(x, y, z)$ coordinates in a given Wyckoff site. To explicitly account for the symmetry, we can augment the representation by considering all possible multiplicities in each Wyckoff site as shown in Figure \ref{fig:rep}c. 

\subsection{Practical Challenges of AI Generative Models for Crystal Structure Prediction}
Using this representation, the objective of CSP is to find or generate the best combination of these variables that leads to a low-energy configuration. Inspired by the successes of AI generative models for image, point cloud, or text, we aim to develop a model to learn the statistical distribution from the training data for both discrete and continuous variables, and then generate new crystal representations. However, there are several challenges that prohibit the direct application of existing AI generative models to crystal generation.

First, there is a risk of lacking sufficient training data to describe the design space for the discrete variables (e.g., space group and Wyckoff site choices). In the context of crystal structure prediction, most materials generation tasks require the researchers to generate likely candidates from only a limited choices of good examples. However, with a total of 230 space groups and 1731 Wyckoff site choices, even for a simple unit cell of 10 atoms, the design space becomes astronomically large. With limited training data in lack of diversity and coverage, it is likely to lead to overfitting and poor generalization, resulting in generated structures that are too similar to the training data or fail to explore the design space effectively. 

Second, assuming the discrete design space (i.e. space group and Wyckoff site choices) has been sufficiently covered, the next challenge is to generate a valid crystal structure under the symmetry constraints. Currently, there exist two common practices to generate the continuous variables: (i) directly applying the diffusion model to generate the continuous variables for unit cell and Wyckoff sites \cite{levy2024symmcd,diffcsp++}; or (ii) combining random sampling with energy-based optimization to search for the low-energy structures \cite{kazeev2025wyckoff, zhu2024wycryst}. The first approach relies on the availability of sufficient training data and may not be applied to small data tasks. Although the second approach is more reliable, for a specific space group and Wyckoff site choice, we need to sample a large number of random configurations and then optimize to find the low-energy states. This is often time-consuming and the quality of the optimized solution is overly reliant on the initial guess.

Third, materials researchers may not only seek target structures optimized for a single metric such as energy or a specific property. In practice, materials are often designed to incorporate established structural motifs or to maintain a preferred local environment (e.g., specific coordination geometry). To our knowledge, existing generative models do not explicitly enforce local geometry during structure generation. As a result, generated structures may deviate from the desired local motifs, limiting their practical relevance for targeted materials design.

\begin{figure*}[htbp]
    \centering
    \includegraphics[width=0.9\textwidth]{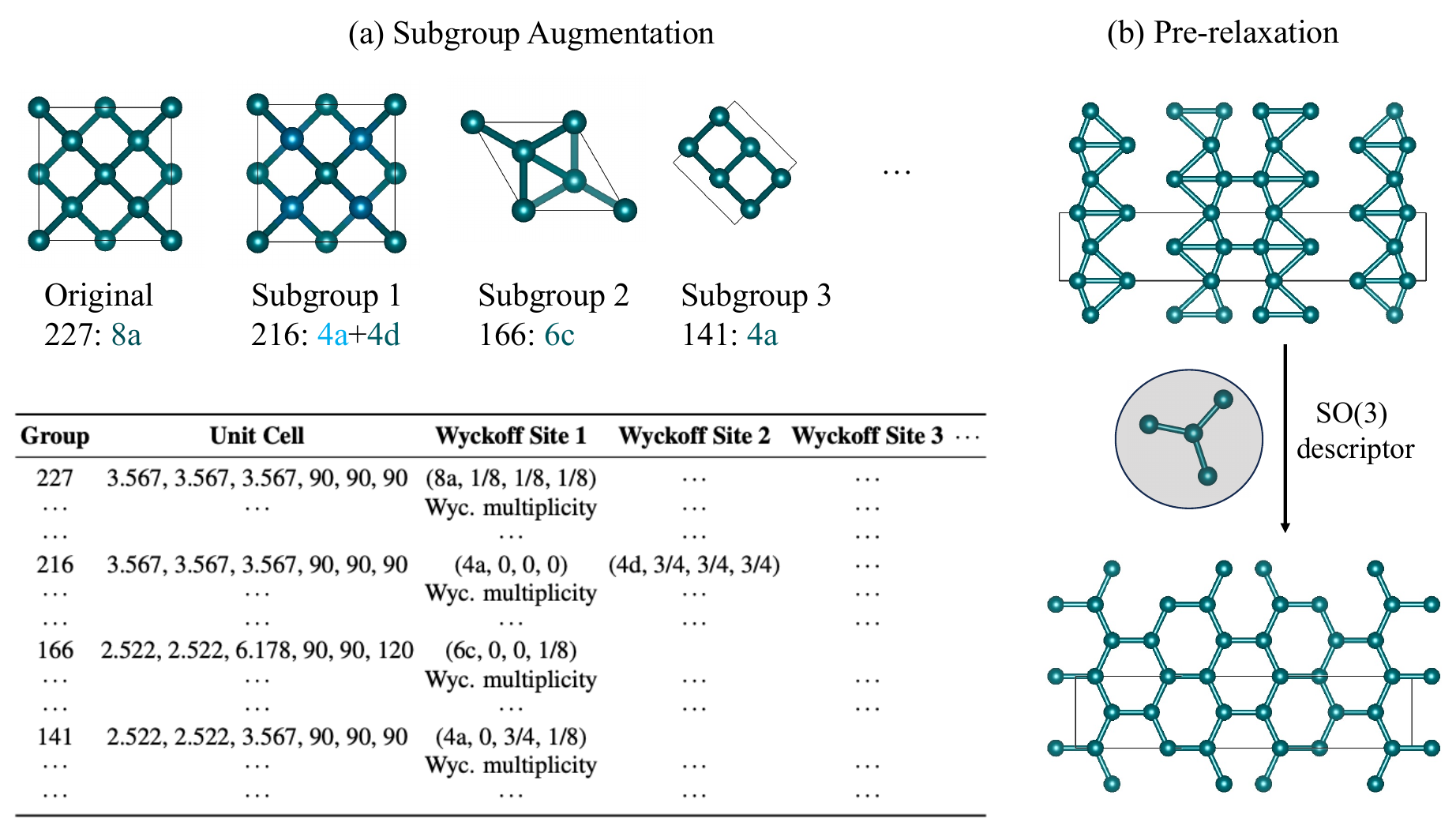}
    \caption{\textbf{Proposed symmetry augmentation and pre-relaxation approaches to enhance the learning of crystal structure representation}. (a) shows the process to augment the representation of diamond crystal by using its associated subgroup symmetry structures. (b) illustrates a graphic example to pre-relax the generated crystal to the desired sp$^2$ local environment by using the power spectrum SO(3) descriptor.}
    \label{fig:aug}
\end{figure*}

\subsection{Symmetry and Geometry Considerations}
To effectively address these fundamental challenges in crystal structure prediction with AI generative models, we propose two key numerical recipes that significantly improve the model's performance as illustrated in Figure \ref{fig:aug}.

\textbf{Data augmentation by subgroup symmetry relations}. By default, a crystal structure is assigned to the highest possible space group symmetry. However, this assignment does not convey the symmetry information comprehensively. For example, the diamond structure can be represented by its highest symmetry space group \textit{Fd}$\bar{3}$\textit{m} (227) with Wyckoff site 8\textit{a}. The cubic boron nitride can be viewed as a split of 8\textit{a} Wyckoff site into two Wyckoff sites (4\textit{a} and 4\textit{d}) in a lower space group symmetry \textit{F}$\bar{4}$3\textit{m} (216). While these two structures are closely related by symmetry breaking, this relationship cannot be recognized by simply comparing their tabular representations using the highest symmetry representation. In fact, such a Wyckoff split is rather common in crystal structures, especially during phase transitions or structural transformations \cite{wondratschek2006symmetry}. Following our recent study on symmetry relations in ferroelectric phase transitions, we leverage our previously developed subgroup symmetry module within the \texttt{PyXtal} library \cite{pyxtal} to generate a rich set of subgroup representations. Using diamond crystal as an example (see Figure \ref{fig:aug}a), we can generate multiple alternative subgroup symmetry representations to provide more training data. This approach enables sophisticated statistical models to thoroughly learn the distribution of space group/Wyckoff combinations in a more comprehensive design space. After training on the symmetry-augmented samples, the model can then generate not just exact replicates of the highest symmetry space groups from the training data, but also closely related subgroup symmetry alternatives.

\textbf{Pre-relaxation with respect to the reference local environment.} Once the model is capable of suggesting a promising candidate crystal, we then need to optimize the remaining continuous variables to get the realistic structure, under the conditioned space group symmetry constraint. In many practical applications, we may already know how the atoms may behave in a short-range cutoff. For instance, carbon may take sp, sp$^2$, or sp$^3$ local bonding; Porous materials such as zeolites and metal–organic frameworks (MOFs) are often categorized by their well-defined molecular building units, making them ideal targets for generative approaches that preserve local chemical environments. Therefore, we can utilize this knowledge to predefine the reference environment according to the recently developed power spectrum descriptors from spherical harmonics expansion \cite{steinhardt1983bond, kazhdan2003rotation, Bartok-PRB-2013, Zagaceta2020SpectralNN, yanxon2020pyxtalff}, and then relax the structure with respect to this reference environment. Figure \ref{fig:aug}b explains the idea of using SO(3) descriptor based reference environment for a randomly generated symmetric crystal.
Ideally, we aim to achieve two outcomes from the use of pre-relaxation, including (1) maintain the same chance if the initial guess is close to an ideal arrangement; and (2) enhance the chance of generating good structures even from a bad guess by pre-relaxation. The numerical details and justification can be found in the supplementary materials Section III. Hence, we can rapidly screen if the trial structure can be relaxed to the desired environment prior to further consideration with more expensive energy-based optimization. Note that similar ideas of optimizing the structures at the descriptor space have also been explored in several recent works \cite{tao2024accelerating, pickard2025beyond}.

Using the aforementioned approaches, we expect that the search space can be greatly reduced for both discrete and continuous variables that are needed to describe the crystal. Therefore, one can manipulate the structural complexity with only a few variables, and the generative model can be trained on a small dataset with a limited number of known examples.

\section{The LEGO-xtal framework}

\begin{figure*}[htbp]
    \centering
    \includegraphics[width=0.95\textwidth]{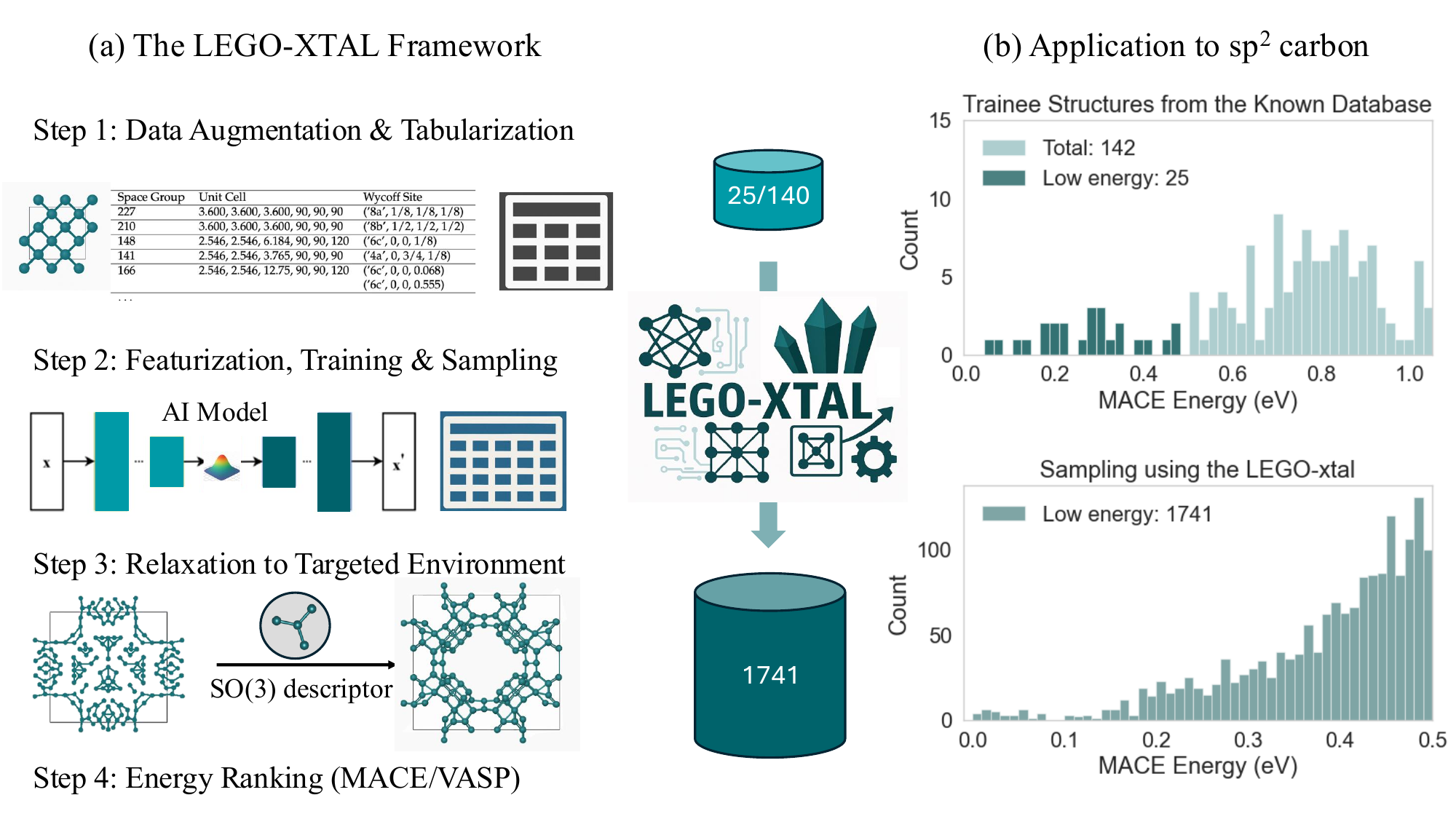}
    \caption{\textbf{The \texttt{LEGO-xtal} framework and its application to search for low-energy carbon allotropes}. (a) shows the detailed pipeline of LEGO-cryst for adaptive generation of complex crystal structures. (b) summarizes the energy distribution of training dataset and newly generated structures by using this framework.}
    \label{fig:overview}
\end{figure*}    

Motivated by preceding discussions, we propose the Local Environment Geometry-Oriented Crystal Generator (\texttt{LEGO-xtal}) framework that is capable of rapidly generating many feasible crystal structures from a few known examples. As illustrated in Figure \ref{fig:overview}, it consists of four main components: (1) data collection and augmentation, (2) generative modeling training and sampling, (3) structural pre-relaxation, and (4) energy ranking with physical models. Below, we will describe the details of each component in the context of generating complex structures comprising exclusively sp$^2$-hybridized carbon atoms. Although the ground state of sp$^2$-carbon is well known to be the graphite structure, many metastable carbon allotropes exist with intriguing properties. The exploration of these sp$^2$ carbon allotropes has been highly sought after due to their fascinating electrical, optical, magnetic, thermal, and chemical properties \cite{hoffmann2016homo, lenosky1992energetics, pun2018toward, braun2018generating}. Hence, we will use the sp$^2$-carbon allotropes as a testbed to demonstrate the effectiveness of the \texttt{LEGO-xtal} framework.

\subsection{Training Data Augmentation \& Tabularization}
First, we need to gather the already known data used for training as many as possible. For carbon research, the online Samara Carbon Allotrope Database (SACADA) database, which contains all available data about 3-periodic carbon allotropes extracted from scientific literature from Web of Science and Scopus databases \cite{hoffmann2016homo}. From SACADA, we extracted 140 carbon allotropes which are made of all sp$^2$ carbon atoms. This includes a total of 127 unique (space group, Wyckoff sites) combinations, with energy spanning from 0 (graphite) to 1.344 eV/atom using the GGA-PBE pseudopotential \cite{PBE-PRL-1996} within the framework of density functional theory (DFT). 

For each structure, we utilized \texttt{PyXtal}'s subgroup symmetry module \cite{pyxtal} to generate valid subgroup representations within constraints of 500 atoms and 8 Wyckoff sites per unit cell. Since the subgroup representation often results in larger unit cells and more Wyckoff sites, we excluded 10 structures with either low symmetry or large atom counts from the original 150 structures (see Supplementary Materials Section IV-A and Table S2). For each of the remaining structures, we generated 2-15 subgroup symmetry representations. Additionally, we randomly sampled the multiplicity of each Wyckoff site due to symmetry operations. To evaluate the effectiveness of data augmentation, we created three datasets labeled as v1-60k, v2-120k, and v3-240k, with details summarized in Table \ref{tab:dataset}.

\begin{table}[htbp]
\centering
\caption{Summary of the training dataset for sp$^2$ carbon allotropes.}
\begin{tabular}{lcc}
\toprule
\textbf{Dataset} & ~~\textbf{Number of samples}~~ & \textbf{Number of unique symmetries\footnotemark[1]}\\
\midrule
Raw       & 140     & 127 \\
v1-60k    & 63,114  & 419 \\
v2-120k   & 122,322 & 449 \\
v3-240k   & 242,766 & 485 \\
\bottomrule
\end{tabular}
\footnotetext[1]{The unique symmetry is defined according to the combination of space group and sorted list of Wyckoff indices. For instance, the diamond carbon is denoted as (227, 8a).}
\label{tab:dataset}
\end{table}

In each representation, the space group number and Wyckoff site column values are always treated as discrete features. The cell parameters and Wyckoff site position values are treated as continuous features by default (\textit{cont-dataset}). To improve the model's learning efficacy, we also explore the possibility of transforming all continuous variables to discrete categorical values. In the discrete mode (\textit{dis-dataset}), we divide the unit cell into a $100 \times 100 \times 100$ grid and convert each Wyckoff position into corresponding grid indices. For the cell parameters, we set up 100 bins between 0 and 35 for cell lengths, and 100 bins between 30 and 150 degrees for cell angles. The number of columns is 39 when the maximum number of Wyckoff sites is 8. Optionally, one may add some new columns, such as energy or other desired physical properties, to inject more information to guide the structure generation process.

\subsection{Generative Model Training \& Sampling}
Prior to employing AI generative models, we transformed each row into an extended feature representation to enhance model learning. For discrete features, one-hot encoding is used, where each unique categorical value is mapped to an individual binary column. In this representation, a value of 1 indicates the presence of the corresponding category, while all other columns are set to 0. For continuous columns, we apply a Bayesian Gaussian Mixture Model (BGMM) approach as proposed by Patki et al \cite{patki2016synthetic}, in which each continuous value is first scaled to the interval $[-1, 1]$, and then softly assigned to distinct Gaussian components via a probabilistic one-hot representation. This dual encoding strategy effectively captures both normalized numerical information and probabilistic cluster associations, improving the stability and efficacy of feature representation during generative model training. In the dis-dataset, we replace the BGMM-based encoding with fixed binning. All continuous features are discretized into categorical bins (e.g., cell lengths into 100 bins between 0–35~\AA{}), and the resulting values are one-hot encoded. This yields a fully categorical representation for the entire dataset.

After transformation, the table expands to 525 columns in the cont-dataset and 2332 columns in the dis-dataset. These two encoding schemes allow us to benchmark model performance under both mixed-type and fully discrete data conditions. Next, we explored two generative models on both dis- and cont-type datasets to learn the underlying distribution of crystal symmetry data. The models are trained to synthesize new rows of tabular features that correspond to novel crystal structures that satisfy crystallographic constraints.

\begin{itemize}
    \item \textbf{GAN}. The Generative Adversarial Network (GAN) framework involves two neural networks—the Generator and the Discriminator—trained in a min-max game fashion, in which the Generator aims to create synthetic samples from random noise and fool the Discriminator, while the Discriminator attempts to distinguish real data from generated data. Our implementation is based on the Conditional Tabular GAN model \cite{ctgan}, which is specifically designed to handle tabular datasets with mixed data types (both continuous and categorical). 
    During training, we adopt the Wasserstein GAN with Gradient Penalty framework \cite{gulrajani2017improved} to improve training stability by enforcing the Lipschitz continuity condition via a gradient penalty.
    Once training is complete, the Discriminator is discarded, and only the trained Generator is used to produce new synthetic samples by using a random noise vectors sampled from a standard normal distribution.
    
    \item \textbf{VAE}. The Variational Autoencoder (VAE) is a probabilistic generative model that learns to map input data into a structured latent space from which new samples can be generated \cite{kingma2013auto, rezende2014stochastic}. The model consists of two neural networks: an Encoder that maps each input to the parameters (mean and variance) of a Gaussian distribution into latent space, and a Decoder that reconstructs the input from a sample drawn from this distribution. The latent space is regularized by enforcing the learned posterior distribution \( q_\phi(z \mid x) \) to be close to a standard normal prior following the standard Normal distribution \( \mathcal{N}(0, I) \), ensuring smoothness and continuity in the latent space which facilitates sampling and interpolation.

    The training objective includes a reconstruction loss and a Kullback-Leibler (KL) divergence term. The reconstruction loss ensures fidelity between the input and output (negative log-likelihood for continuous features, cross-entropy for categorical ones), while the KL term encourages the approximate posterior to match the prior, acting as a form of regularization. During the inference phase, new samples are generated by drawing random noise vectors from the standard multivariate normal distribution. These latent vectors are passed through the trained Decoder network, which maps them to the transformed tabular representation of crystal symmetry features. 
\end{itemize}

For more details regarding GAN and VAE models, please refer to the Supplementary Materials Section I and Figs. S1 and S2). In a typical experiment, we train each GAN or VAE model with 250-1000 epochs, which is sufficient to converge the loss function values. Then we take the epoch that achieves the lowest loss function value to generate 100k - 500k samples for further consideration. 

\subsection{Structure Relaxation to Targeted Local Environment}
Next, the structures yielded from the trained models need to be relaxed to the nearby energy minimum with the desired geometries. Instead of using the conventional structure optimization techniques based on classical force fields or electronic structure theory, we employed a descriptor-guided optimization framework that minimizes the dissimilarity in local atomic environments relative to a reference structure. 

To represent the local atomic environment, the SO(3) descriptor $P$, with the option to include the explicit radial distribution function, is calculated to extract the radial and angular features of the neighbor density function \cite{yanxon2020pyxtalff}.
For the purpose of this work, we selected a sp$^2$ carbon (i.e., one centered atom is connected to 3 neighboring carbon with 1.42~\AA~ bond length and 120$^\circ$ bond angles in the same plane) to compute its power spectrum $P_{\text{ref.}}$ up to a radial cutoff of 2.1 \AA. For each generated structure, we first evaluated the corresponding power spectrum $P$ for each atom, and then sought to minimize the summed mean squared error (MSE) loss function as follows:

\begin{equation}
F_\text{objective} = \sum^{\text{all atoms}}_i\|P_i - P_{\text{ref.}}\|^2,
\end{equation}
which quantifies the structural deviation from the standard sp$^2$ environment. The goal of the optimization procedure is to adjust reduced variables (e.g., lattice parameters and free $x, y, z$ Wyckoff positions) to minimize this MSE loss function value, thereby steering structures toward graphite-like local order without requiring explicit energy evaluations.
To perform the minimization, we utilized available optimization algorithms from the \texttt{scipy.optimize.minimize} library \cite{scipy}, including Nelder-Mead \cite{nelder1965simplex}, L-BFGS-B \cite{zhu1997algorithm}, and Basin-Hopping \cite{Wales-JPCA-1997}.

During optimization, symmetry constraints were naturally enforced (due to the use of reduced variables) to ensure chemical realism and structural validity.
Combining with the subgroup representation, this descriptor-guided strategy not only shortens the computational overhead associated with iterative energy evaluations but also allows the system to escape undesirable local minima associated with conventional force-based optimization. While we may use the subgroup symmetry as the initial guess, it is important to note that the low-symmetry structure after relaxation does not necessarily evolve to high symmetry. This is because many pathways may exist to stabilize the Wyckoff site to a target reference environment.
For more details, please refer to Fig. S3 and a motivating example in the Supplementary Section III. 

\subsection{Energy Ranking \& Database}
After the pre-relaxation, the valid structures need to be evaluated by the energy ranking. Given that many structures needs to be processed and it is computationally expensive to perform quantum mechanical calculations for each structure, we considered two alternatives, (i) a reactive force field (ReaxFF) \cite{van2001reaxff} via the \texttt{GULP} package \cite{gulp}, a bond-order-based force field that can efficiently model chemical reactions and interatomic interactions; and (ii) a pretrained MACE model \cite{Batatia2022mace} via \texttt{ASE} \cite{ASE}, which utilizes higher order equivariant message passing for fast and accurate predictions of interatomic interactions in a large chemical space. Both methods are significantly faster than conventional DFT calculations, making them suitable for large-scale screening of generated structures. Although most of the pre-relaxed structures should be very close to the ground state, each energy model may describe their ground state slightly differently in terms of bond lengths and angles, as well as lattice constants. Therefore, it is still necessary to run relaxation, instead of single point energy evaluation, to obtain the ground state and accurate energy ranking.

\begin{figure}[htbp]
    \centering
    \includegraphics[width=0.45\textwidth]{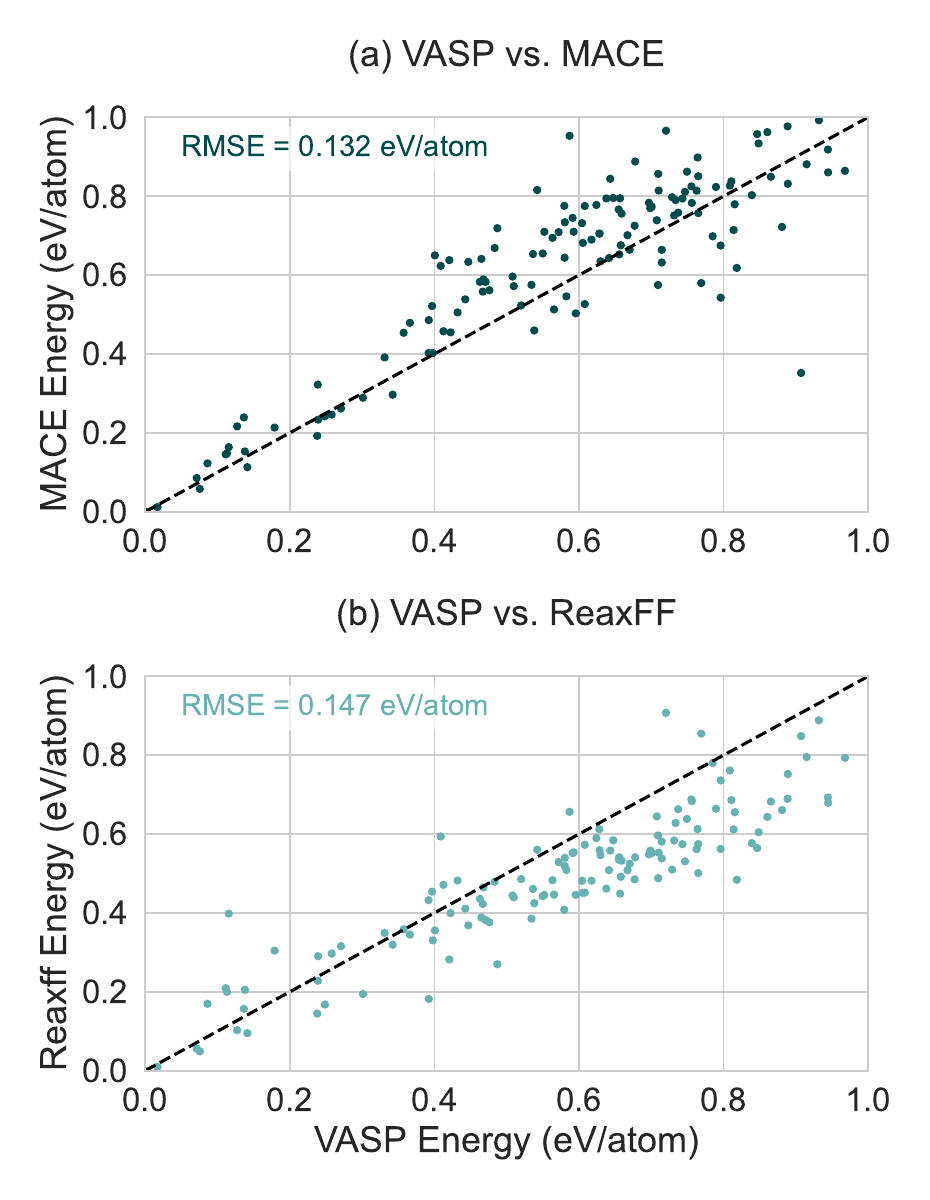}
    \caption{\textbf{Comparison of simulated energies for the selected 140 sp$^2$ allotropes from the SACADA database.}. (a) VASP vs. MACE and (b) VASP vs. ReaxFF. The dashed line indicates the ideal 1:1 correspondence.}
    \label{fig:energy}
\end{figure}

Figure \ref{fig:energy} compares energies calculated using MACE and ReaxFF models against reference DFT calculations performed with the \texttt{VASP} code \cite{Vasp-PRB-1996} using the r$^2$SCAN meta-GGA functional \cite{furness2020accurate, furness2020correction} and a KSPACING value of 0.15 \AA$^{-1}$. In terms of the the root mean square error (RMSE) values, the MACE model (0.132 eV/atom) shows a better agreement with DFT energies compared to ReaxFF (0.147 eV/atom). Particularly, MACE describes better for the low-energy structures, making it slightly more suitable for our purposes. However, both models may stabilize structures into nonphysical configurations after pre-relaxation. Therefore, cross-checking generated structures with multiple models helps ensure physical validity. In our pipeline, we first use ReaxFF for relaxation and energy ranking, then apply MACE optimization to get the energy value as an alternative metric to infer the structural stability. According to our empirical observation, a feasible structure should be characterized by favorable energy ranking in both MACE and ReaxFF models.

The final database includes structures with MACE energies below 0.5-1.0 eV/atom. Beyond energy calculations, we characterize structural topology using \texttt{CrystalNet.jl} \cite{zoubritzky2022crystalnets} to identify novel configurations compared to existing databases using the common RCSR notations \cite{rcsr}, providing additional insight into the structural diversity and stability of generated structures.

\subsection{Approximated Time Costs}
Table \ref{tab:time} summarizes the approximate timing for each stage in the \texttt{LEGO-xtal} workflow. All timings are based on processing 1,000 samples; for larger datasets, times scale linearly.

First, Data augmentation is performed on a single CPU and typically takes 0.2–0.5 minutes per 1,000 samples. Model training on an H100 GPU requires 10–20 minutes for 250 epochs, while sampling (generation of new structures) takes 1–2 minutes per 1,000 samples.

Next, Pre-relaxation using the SO(3) descriptor on an AMD EPYC 96-core processor (2.4 GHz) takes approximately 5.5 minutes per 1,000 samples. When SO(3) pre-relaxation is followed by GULP-ReaxFF or MACE, the combined times are 7.5 and 9.5 minutes per 1,000 samples, respectively. For comparison, running GULP-ReaxFF or MACE without pre-relaxation requires 20 and 120 minutes per 1,000 samples, respectively. This suggests that pre-relaxation can effectively reduce the time cost as compared to the energy-based optimization. As shown in the Supplementary Materials Section III-C, we also provide direct evidence to demonstrate that pre-relaxation also improve the success rate of finding the unique valid structures.

In addition, we have implemented Batchwise SO(3) pre-relaxation on an H100 GPU which is even faster, taking about 1.5 minute per 1,000 samples, although this approach is still under development and requires further optimization.

\begin{table}[htbp]
\centering
\caption{Summary of time costs for each stage in \texttt{LEGO-xtal}.}\label{tab:time}
\begin{tabular}{llr}
\toprule
Stage & Device & Speed (per 1k samples) \\
\midrule
Data Augmentation   & Single CPU    &  0.2-0.5 mins \\
Training     & H100 GPU   &  10\textasciitilde20 mins / 250 epochs \\
Sampling            & H100 GPU  &  1\textasciitilde2 mins \\
Pre-relaxation      & AMD 96-core  &  \textasciitilde ~5.5 mins \\
Pre-relax + ReaxFF & AMD 96-core  &  \textasciitilde ~5.5+1.5 mins\\
Pre-relax + MACE & AMD 96-core  &  \textasciitilde ~5.5+2.5 mins \\
Pre-relaxation Batchwise    & H100 GPU &  \textasciitilde1.5 mins  \\
ReaxFF w/o Pre-relax   & AMD 96-core  &  \textasciitilde 20 mins \\
MACE w/o Pre-relax   & AMD 96-core  &  \textasciitilde 120 mins \\
\bottomrule
\end{tabular}
\end{table}

\section{Results \& Discussions}
To evaluate our framework, we trained multiple GAN and VAE models for generating sp$^2$ carbon allotropes, by experimenting with various hyperparameters (e.g., training dataset size, continuous versus discrete representation, neural network architectures, number of training epochs). From each model, we generated 200k samples for evaluation. After pre-relaxation, we only retained structures exhibiting valid sp$^2$ environments for further processing using MACE and ReaxFF. Below, we first analyze how different hyperparameter choices impact model performance, followed by a detailed examination of the sp$^2$ allotrope database generated using \texttt{LEGO-xtal}.

\subsection{Model Performances Evaluation}
Table~\ref{tab:overlap_sp2} shows statistics of unique sp$^2$ structures from different generative models trained on the v1-60k dataset with 200\textasciitilde500 epochs. Compared to the original 140 unique structures in the SACADA dataset, all models except GAN-Cont. successfully regenerated over 100 training structures within 100k samples, indicating a strong capability to learn from the training data. If we allow generating more samples, we expect all the training structures would be reproduced by these models. More encouragingly, each model generated a substantial number of unique structures (GAN-Cont.: 2937, GAN-Dis.: 4514,
VAE-Cont.: 4862, VAE-Dis.: 4289). The limited overlap between models, as found in Table \ref{tab:overlap_sp2}, suggests that each model learn different statistical patterns from the training data, leading to diverse generated structures.

\begin{table}[htbp]
\centering
\caption{The overlap matrix showing the number of shared unique sp$^2$ structures between 140 training structures and the 100k samples trained from v1-60k dataset using different models.}
\label{tab:overlap_sp2}
\begin{tabular}{lccccc}
\toprule
\textbf{} & \textbf{Train} & \textbf{GAN-Cont.} & \textbf{GAN-Dis.} & \textbf{VAE-cont.} & \textbf{VAE-Dis.} \\
\midrule
\textbf{Train} & 140 & 62 & 111  & 100 & 132\\
\textbf{GAN-Cont.}    &  & 2937 & 649 & 848 & 707 \\
\textbf{GAN-Dis.}     &  & & 4514 & 979 & 952 \\
\textbf{VAE-Cont.}    & & & & 4862 & 1152 \\
\textbf{VAE-Dis.}     & & & & & 4289 \\
\bottomrule
\end{tabular}
\end{table}

\begin{table*}[htbp]
\centering
\caption{Summary of metrics evaluated on 100k generated structures across different models with the best numerical results highlighted bold.}\label{tab_summary}
\renewcommand{\arraystretch}{1.3}
\setlength{\tabcolsep}{4pt} 
\begin{tabular}{llrrrrrr}
\toprule
\textbf{Model}~~~~ & \textbf{Dataset}~~~~ & \textbf{$N_{\text{valid\_xtal}}$} & \textbf{$N_{\text{valid\_env}}$} & \textbf{$N_{\text{unique}}$} & \textbf{$N_{\text{train}}$} & \textbf{$N_{\text{lowE\_all}}$} & \textbf{$N_{\text{lowE\_cubic}}$} \\
\midrule
\multirow{3}{*}{GAN-Cont.} 
& v1-60k & 95,313 & 10,687 & 2,937 & 62  & 147 & 30 \\
& v2-120k & 96,174 & 10,486 & 3,132 & 65  & 40  & 31 \\
& v3-240k & 96,404 & 9,621 & 3,235 & 71  & 43  & 29 \\

\midrule
\multirow{3}{*}{GAN-Dis.} 
& v1-60k & 98,393 & 16,985 & 4,514 & 111 & 212 & 62 \\
& v2-120k & 98,861 & 18,406 & 4,814 & 120 & \textbf{426} & 69 \\
& v3-240k & \textbf{99,026} & 16,092 & 5,080 & 119 & 291 & \textbf{70} \\

\midrule
\multirow{3}{*}{VAE-Cont.} 
& v1-60k & 97,657 & 16,524 & 4,862 & 100 & 57  & 35 \\
& v2-120k & 97,812 & 15,330 & 5,123 & 94 & 206 & 28 \\
& v3-240k & 98,004 & 15,052 & \textbf{5,156} & 101 & 241 & 28 \\

\midrule
\multirow{3}{*}{VAE-Dis.} 
& v1-60k & 97,466 & \textbf{22,204} & 4,640 & \textbf{133} & 267 & 59 \\
& v2-120k & 97,419 & 15,685 & 4,649 & 127 & 53  & 36 \\
& v3-240k & 98,302 & 15,123 & 4,883 & 118 & 51  & 35 \\
\bottomrule
\end{tabular}
\end{table*}

Next, we performed a more extensive benchmark to assess the quality and success of our generative method with different hyperparameters. To make a meaningful comparison between different models, we introduced the following metrics:
\begin{itemize}
    \item $N_{\text{valid\_xtal}}$: Number of generated samples that match the required constraints of the intended crystal symmetry. 
    \item $N_{\text{valid\_env}}$: Number of valid structures successfully optimized to adopt the sp$^2$-type bonding configuration.
    \item $N_{\text{unique}}$: Number of unique crystal configurations among the valid set, after removing duplicates.
    \item $N_{\text{train}}$: Number of structures that were regenerated from the initial 140 training set. 
    \item $N_{\text{lowE\_all}}$ and $N_{\text{lowE\_cubic}}$: Number of unique (cubic) crystals with a MACE energy less than 0.55 eV/atom compared to the ground state, indicating thermodynamic stability. 
\end{itemize}

These quantitative indicators collectively provide insights into how realistic, novel, and stable the generated structures are. The results, summarized in Table~\ref{tab_summary}, show the relative performance of different models. 
Among them, the GAN models trained on continuous representation appear to perform notably worse than other models in nearly all metrics. This is likely due to the GAN's difficulty in capturing the discrete nature of crystal symmetries and Wyckoff sites, which are crucial for generating valid crystal structures.

Nevertheless, other VAE and GAN models demonstrate similar performances across various metrics. On average, we observe that around 98\% success rate in generating valid crystal structures that satisfy the space group and Wyckoff site symmetry constraints, and 15-22\% of these structures can be relaxed to adopt the desired sp$^2$ bonding configuration. The number of unique structures ranges from 3,235 to 5,156 per 100k samples, indicating a good diversity in the generated crystal configurations. As shown in Fig. S4 of the Supplementary Materials, one can continue to try more samples to get new unique structures. In addition, our analysis in Section V-A of Supplementary Materials suggests that these metrics are clearly better than the pure random sampling approaches, suggesting the efficacy of the statistical model learning.

For realistic application, the more important metric is the number of low-energy unique structures. 
The GAN-Dis-v2/v3 model achieved higher numbers of low-energy (cubic) structures as compared to the corresponding v1 model, suggesting that more explicit subgroup symmetries can promote the model's learning efficacy. A similar trend is also observed for the VAE-continuous models. However, this is not the case for the VAE-Discrete models, in which the model trained on the v1-60k dataset produced the highest number of low-energy structures. This may be due to the fact that the VAE-Dis. model uses the noise sampled from continuous Gaussian distribution, which may hinder the decoder from learning discrete, one‑hot‑like representations in the latent space, thus reducing generation fidelity on categorical outputs. A natural remedy is to adopt the Gumbel‑Softmax reparameterization trick, which introduces a temperature‑controlled, differentiable approximation to categorical sampling: starting from a high temperature to ensure smooth gradients, and annealing towards a low but non‑zero $\tau$ to encourage near‑discrete latent codes while preserving differentiability \cite{jang2017categorical}. Additionally, balancing the KL-divergence and reconstruction terms via $\beta$-VAE \cite{higgins2017beta} or KL-annealing \cite{fu2019cyclical} strategies can prevent posterior collapse and improve latent expressiveness. Finally, in scenarios with severely imbalanced categorical outputs, applying class-weighted Cross-Entropy or Focal Loss \cite{lin2017focal} can stabilize training by emphasizing hard or rare classes. 

Unlike VAE models that explicitly reconstruct inputs, GANs rely solely on adversarial feedback. This enables the discriminator to learn the joint distribution of the representations without being constrained to memorizing input patterns, thereby enhancing the generator’s ability to generalize. This effect is particularly evident in Table \ref{tab_summary}, where the GAN achieves stronger performance on the discrete dis-dataset compared to the Cont dataset. The smaller and more structured joint distribution space of discrete features makes it easier for the GAN to capture meaningful generalizable patterns, rather than overfitting to training samples.

While it is difficult to make a definitive recommendation on a single best model for this task, we observe a general trend that most models can rapidly generate a large number of low-energy sp$^2$ carbon allotropes with diverse structures, as compared to the original 140 training structures. Importantly, Table \ref{tab_summary} suggests that the more data is included, the chance of generating unique structures ($N_{\text{unique}}$) becomes higher regardless choice of generative models. This is likely to be associated with the increment of unique symmetries (see extended discussion in Section V-B of the Supplementary Materials). Overall, the results highlight new opportunities for effectively harnessing the power of AI generative models to explore the structural space of complicated materials. The improvements of model performance will be left for the future, as it is beyond the scope of this work.

\subsection{The sp$^2$ Carbon Allotrope Database}

\begin{figure*}[htbp]
    \centering
    \includegraphics[width=0.99\textwidth]{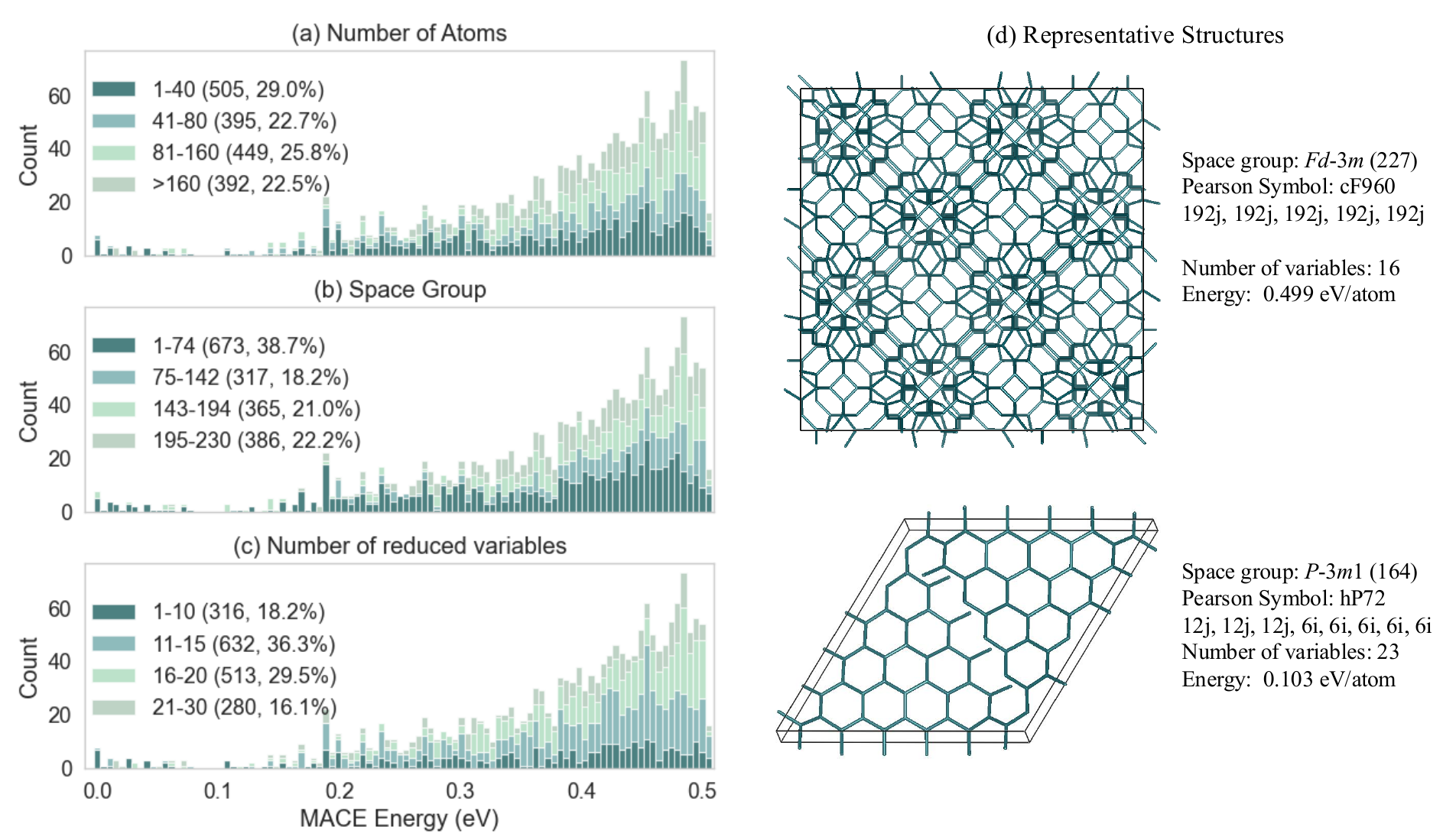}
    \caption{\textbf{Complexity analysis of the 1741 low-energy sp$^2$ carbon allotropes}. (a)-(c) shows the distributions of the energy of the generated structures by number of atoms in the unit cell, space group symmetry, and number of reduced variables, respectively. (d) shows two representative low-energy structures demonstrating the intrinsic crystal searching challenge. In (d), the structures are labeled by their Pearson symbol and relative DFT energies as compared to the ground state graphite structure. The Pearson symbol is a shorthand notation that describes the crystal structure, including the space group and the number of atoms in the unit cell.}
    \label{fig:dist}
\end{figure*}

Since no single model significantly outperformed the others, we ran multiple models to comprehensively explore the structural space of sp$^2$ allotropes. After removing duplicate structures, we identified a total of 21,786 structures within a 1.0 eV MACE energy window, including 1,741 structures with energies less than 0.5 eV above the ground state. 
In the original 140 data used in training, there exist 25 sp$^2$ carbon allotropes that have the MACE energy less than 0.5 eV/atom higher than the ground state graphite structure (see Figure \ref{fig:overview}b). Using the \texttt{LEGO-xtal}, we have significantly expanded it to 1700+ distinct structures, evidently demonstrating the power of our \texttt{LEGO-xtal} framework to rapidly generate novel structures. To double check the accuracy of MACE energy ranking, we also manually selected 150 structures for further relaxation with DFT-r$^2$SCAN using the \texttt{VASP} package. 
The complete database of 1,700+ low-energy sp$^2$ carbon allotropes, as well as ReaxFF, MACE and DFT-r2SCAN energies, is available at \url{https://lego-crystal.onrender.com}.
Below, we will focus on the chemical analysis of these low-energy structures. 

\subsubsection{Distribution of low-energy crystals by complexity}

Figure \ref{fig:dist} displays the distribution of structures by number of atoms per unit cell, space group symmetry, and number of reduced crystal variables in each allotrope. First, Figures \ref{fig:dist}a and \ref{fig:dist}b list the distributions by unit cell size and space group symmetry. Clearly, there exists a large percentage of structures having both high symmetry and large unit cell size. The largest low-energy sp$^2$ structure we attempted has 960 atoms in the unit cell with \textit{Fd}$\bar{3}$\textit{m} space group symmetry, consisting of 5 general Wyckoff sites at 192\textit{i}. To our knowledge, such a large structure has never been reported for either a traditional or AI-based crystal structure prediction method, suggesting that our new approach can efficiently handle the complex structures by taking advantage of crystallographic symmetry. Indeed, as long as the space group symmetry and Wyckoff choices are known, this 960-atom cubic structure contains only 16 reduced crystal variables (1 for the lattice parameter and 15 for ($x, y, z$) coordinates of the five Wyckoff sites). Therefore, the searching space can be drastically reduced. In turn, our approach can effectively generate large crystals in high symmetry, which is a significant advance as compared to recently proposed approaches that are mostly suitable for structures with no more than 20 atoms in the unit cell \cite{zeni2025generative, DiffCSP, merchant2023scaling}. 

In addition, it is important to emphasize that our approach does not necessarily only favor high-symmetry structures. As shown in Figure \ref{fig:dist}b, there exist many structures with lower symmetry space groups other than the cubic symmetry. Using the number of reduced crystal variables as a metric to probe the model's predictive capabilities (Figure \ref{fig:dist}c), we find that the majority of low-energy structures have fewer than 20 reduced variables. This indicates that our framework is most effective at exploring the structural space of sp$^2$ carbon allotropes with a limited number of variables, less than 20. However, we can still find that a notable portion of structures (280 out of 1741) have more than 20 reduced variables. Figure \ref{fig:dist}d highlights a representative low-energy structure with 23 reduced variables - a 3D-periodic structure containing 72 atoms per unit cell in $P\bar{3}m1$ symmetry. This finding demonstrates that our framework can effectively explore more complex structural spaces despite increasing uncertainty at higher dimensions.

\begin{figure*}[htbp]
    \centering
    \includegraphics[width=0.9\textwidth]{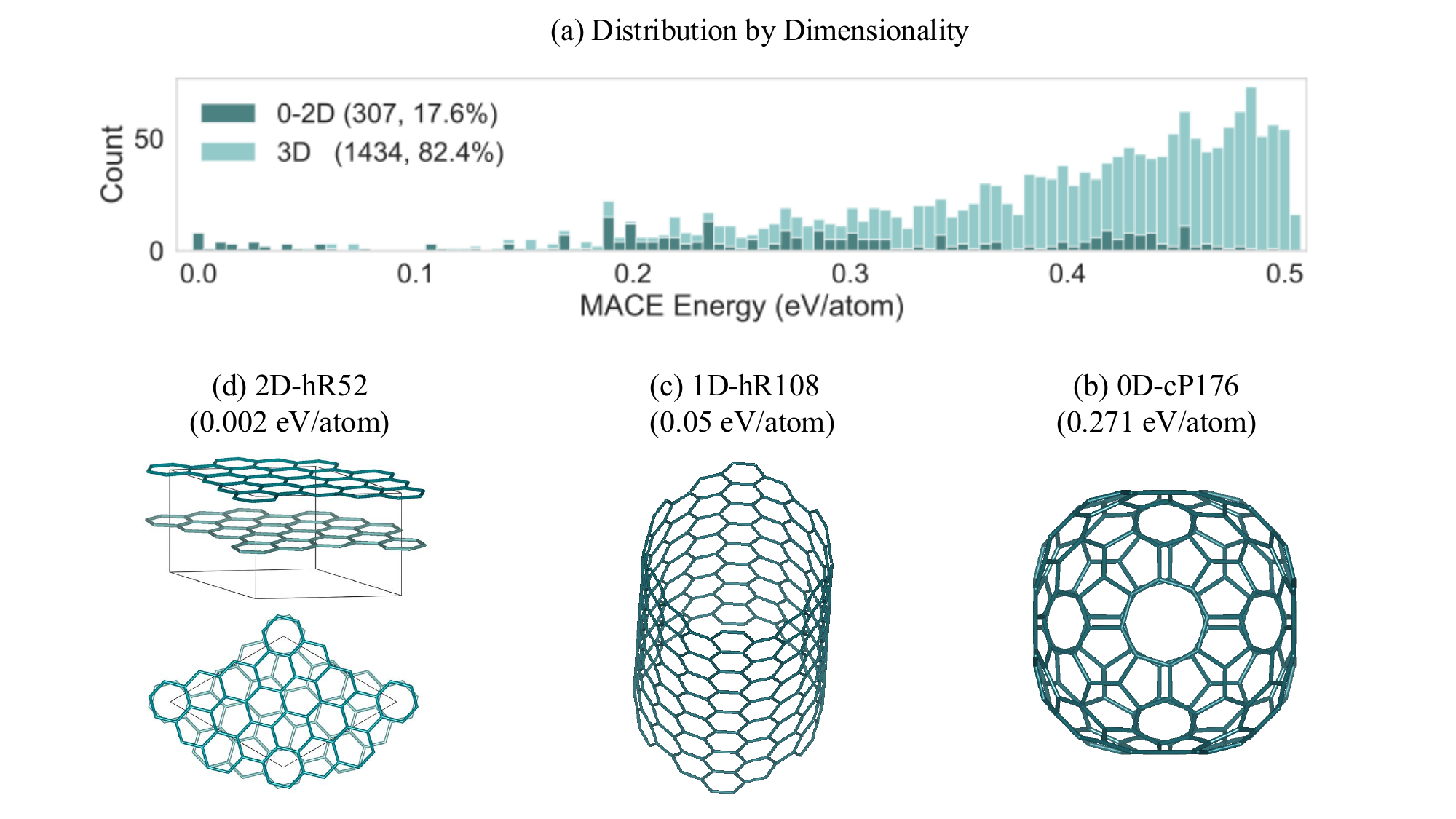}
    \caption{\textbf{The screened low-dimensional sp$^2$ carbon allotropes.} (a) Distribution of the energy of the generated structures grouped by dimensionality. (b-d) show three representative 0D, 1D, and 2D structures labeled by the Pearson symbol and the relative DFT-r$^2$SCAN energies as compared to the ground state graphite structure.}
    \label{fig:lowD}
\end{figure*}

\subsubsection{The 0-2D low-energy sp\texorpdfstring{$^2$}{2}  carbon allotropes}
Given the nature of sp$^2$ bonding, layered graphite is unquestionably the ground state. Figure \ref{fig:lowD}a shows the breakdown of energy distribution by detected dimensionality. The majority of low-energy structures (< 0.15 eV/atom) are characterized by stacked hexagonal graphite layers. In addition to the commonly known stacking polytypism due to layer translations \cite{QZhu-PRB-2015}, we also discovered many variations arising from layer twisting. Among these, the most interesting structure is shown in Figure \ref{fig:lowD}d, which is an AB-type stacked layered structure with about a 30-degree rotation angle between two adjacent layers. This structure is not only low in energy but also exhibits a unique 2D periodicity. This finding suggests that our framework can effectively explore the structural space of sp$^2$ carbon allotropes that is related to recent work on the twisted bilayer graphene structures with varying rotation angles \cite{bistritzer2011moire, cao2018unconventional}.

Beyond the stacking graphite-type layers, we also identified several other planar configurations in different types of ring topology (e.g., mixed five-, seven, eight-membered rings, see Section VI and Fig. S6 in the Supplementary Materials) with an energy spanning from 0 to 0.5 eV range. Additionally, several unique 1D carbon nanotube-like structures with varied diameters (e.g., a representative model 1D-hR108 in Figure \ref{fig:lowD}c) have been observed in our results. Finally, we also found two large 0D allotropes (see 0D-cP176 in Figure \ref{fig:lowD}d) featured by the combination of five, six and eight-membered rings that are similar to the well known C$_{60}$ and C$_{70}$ \cite{c60}. Notably, these 0D allotropes are calculated to possess lower energies (0.2-0.3 eV/atom) than C$_{60}$ and C$_{70}$. Hence, they have better stability if they can be synthesized in the experiment. These examples demonstrate our framework's capability to explore diverse structural motifs across multiple dimensionalities.

\begin{figure*}[htbp]
    \centering
\includegraphics[width=0.95\textwidth]{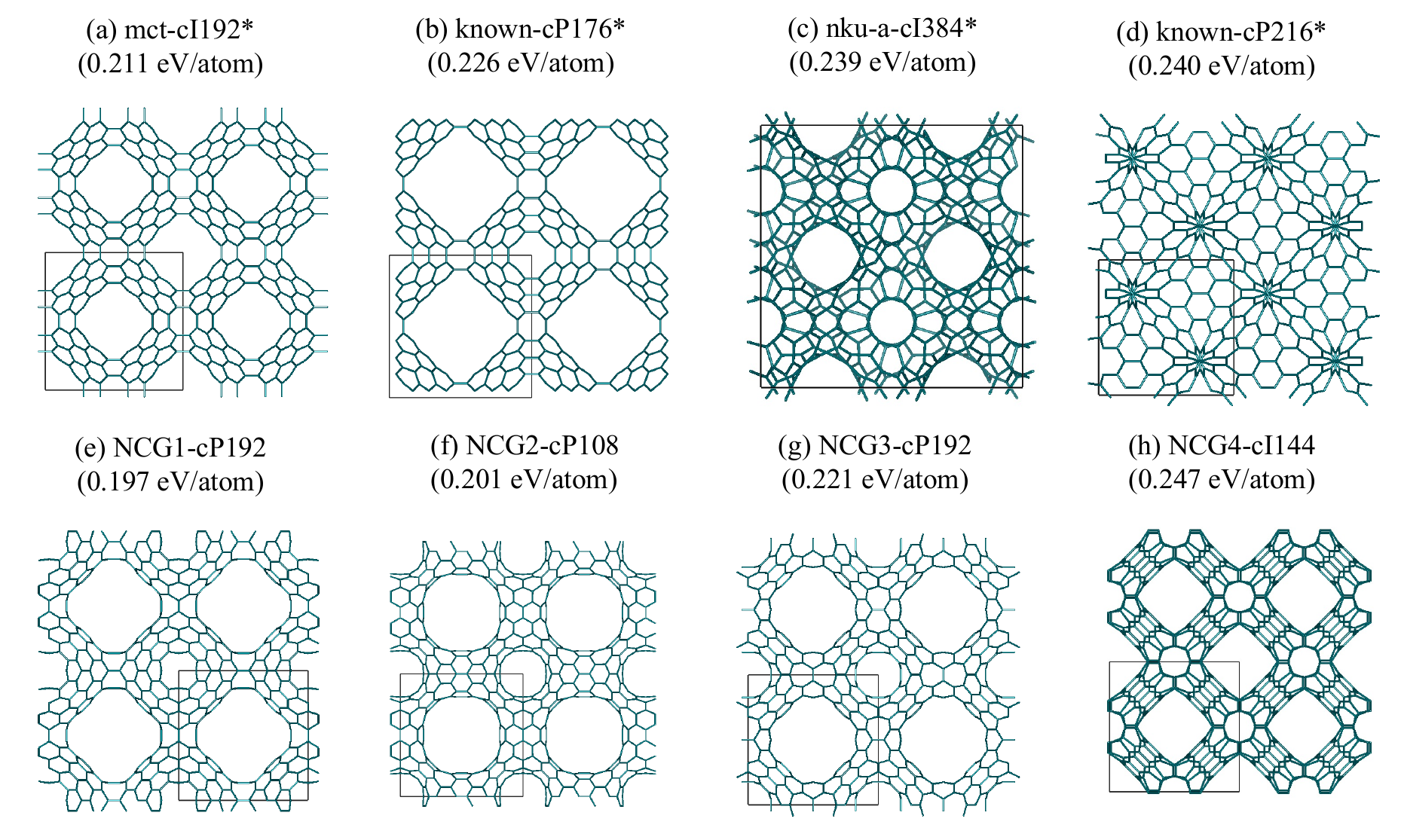}
    \caption{\textbf{A list of low-energy negative curved graphite structures found by \texttt{LEGO-xtal}.} The structures are labeled by the topology according to RCSR notation \cite{rcsr} (when available), the Pearson symbol, and relative DFT-r$^2$SCAN energies as compared to the ground state graphite structure. The previously reported structures (a-d) are also marked by the asterisks.}
    \label{fig:NCG}
\end{figure*}

\subsubsection{The 3D Negative Curved Graphite allotropes}
Apart from the 300+ low-dimensional crystal packing motifs, the majority of low-energy structures possess extended networks with 3D periodicity. Not surprisingly, the lowest energy 3D sp$^2$ allotropes closely resemble graphite-like layered structures (see Fig. S7 in the Supplementary Materials and previous literature \cite{zhao2020family}). However, the conceptually more compelling 3D sp$^2$ allotropes are those containing negative Gaussian curvature that form inward curved surfaces, resembling saddle shapes or hyperbolic geometry \cite{pun2018toward}. Unlike C$_{60}$ with a positive curved graphite network, these 3D negative curved graphite (NCG) structures are more complex and feature fascinating properties due to their unique topology. Since 1990, only 15 NCG structures have been proposed through either mathematical derivation or sophisticated physical modeling approaches \cite{mackay1991diamond, Keeffe1992, lenosky1992energetics, mackay1993hypothetical, mct, mct1, tagami2014negatively, braun2018generating}.

Within the 1741 structures, we have observed a total of 386 sp$^2$ cubic structures. To confirm their stability, we selected the 100 NCG candidates for further energy ranking using \texttt{VASP} at the DFT level. And the top 8 ranked structures are listed in Figure \ref{fig:NCG}. Among them, we found four new structures (NCG1-cP192, NCG2-cP108, NCG3-cP192, and NCG4-cI144) that have neither been used in training nor reported in the previous literature. 
Remarkably, both NCG1-cP192 and NCG2-cP108 are $\sim$0.015 eV/atom lower in energy than the previously known lowest-energy mct-cI192 NCG structure \cite{mct1}. Furthermore, we computed the phonon dispersions for the NCG1-NCG3 structures and confirmed their dynamical stability (see Fig. S8 in the Supplementary Materials).

In addition, there exist 41 structures with more than 300 atoms in the unit cell, highlighting that our approach can effectively handle complex structures that were seldom reported in the literature. Interestingly, by comparing with the structures used in training, we found that all newly identified structures exhibit some similar features to the training data while still maintaining unique characteristics. This indicates that our framework can effectively explore and expand upon known structural motifs rather than simply replicating existing structures. These remarkable findings demonstrate the power of our framework to discover new materials with complex geometries.

\section{Conclusions}
In this work, we present \texttt{LEGO-xtal}, a novel framework that combines subgroup symmetry augmentation, descriptor-guided pre-relaxation, and modern generative components (GAN and VAE) to efficiently explore crystal structure space from limited training data. Starting from just 25 (140) known sp$^2$ carbon structures, our approach successfully generated over 1,700 (21,786) diverse low-energy allotropes, including high-symmetry structures with large unit cells (up to 960 atoms) and unique 0D, 1D, and 2D configurations. This demonstrates the framework's capability to discover complex crystal structures from a small dataset while maintaining physically desirable local environments. This clearly distinguishes our approach from previous methods \cite{zhu2024wycryst, merchant2023scaling, CDVAE, DiffCSP, zeni2025generative, levy2024symmcd, diffcsp++, miller2024flowmm, zhong2025practical, antunes2024crystal, gruver2024fine, cao2024space, kazeev2025wyckoff} that focus on generation of relatively small crystal systems from a large dataset without the explicit constraints on the local chemical environment.

The success of our framework stems from the synergistic combination of subgroup symmetry augmentation and descriptor-guided pre-relaxation. While symmetry augmentation enables efficient exploration of the structural space through reduced discrete variables, pre-relaxation ensures the generated structures maintain desired local environments at minimal computational cost. Together, these strategies dramatically reduce the dimensionality of the crystal structure search space. Importantly, the subgroup-based data augmentation strategy introduced in our work is broadly applicable. In the supplementary materials Section IV-C, we have provided a universal interface to systemically search for subgroup symmetries, which could be integrated into existing symmetry-aware generative models to enhance their generalization capabilities \cite{zhu2024wycryst, levy2024symmcd, diffcsp++, antunes2024crystal, cao2024space, kazeev2025wyckoff}.

Regarding the choice of generative models, this study explored both VAE and GAN architectures that offer a balance between simplicity and effectiveness. Despite their relative simplicity, both remain core building blocks in today’s state-of-the-art pipelines (e.g., Stable Diffusion \cite{rombach2022high}, SV4D 2.0 \cite{yao2025sv4d}, MaskGIT \cite{chang2022maskgit}). We hope that LEGO-xtal will serve as a solid starting point, from which we can build toward integrating more advanced generative modeling techniques more advanced paradigms such as latent diffusion, masked modeling, or auto-regressive architectures in future work.

Due to the use of symmetrized tabular representation, this approach is likely to intrinsically favor high symmetry structures. For instance, our current models do not consider the generation of more than 8 atoms in the unit cell with $P$1 symmetry. While the assumption of no low-energy complex structures in low symmetry is true for the case of sp$^2$ carbon allotropes, it may not be suitable to describe other systems where low-symmetry crystals appear more often. In that case, we recommend the use of the generative models without symmetry constraints \cite{zeni2025generative, DiffCSP, miller2024flowmm}.

Finally, we aim to extend subgroup symmetry augmentation beyond single components and reference environments, enabling the generation of more complex structures with multiple components and varied local environments in a single model. This extension requires (1) adapting pre-relaxation strategies to handle multiple reference environments, either by modifying the environmental loss function within traditional optimization routines \cite{pickard2025beyond} or by employing more flexible diffusion models that incorporate local environment constraints during the training stage and (2) the support of generative models on Wyckoff sites labeled with different elements or local environments. Conceptually, the use of descriptor-guided pre-relaxation and subgroup augmentation is expected to be even more advantageous to handle more complex compositional systems. These improvements will enable the generation of increasingly complex structures with higher dimensionality and more sophisticated local environments, which will be the focus of our future work.

\section*{Acknowledgments}
This research was sponsored by the U.S. Department of Energy, Office of Science, Office of Basic Energy Sciences, and the Established Program to Stimulate Competitive Research (EPSCoR) under the DOE Early Career Award No. DE-SC0024866, the UNC Charlotte's seed grant for data science, as well as European Union (ERDF), Région Nouvelle Aquitaine, Poitiers Univeristy, and French government programs ``Investissements d'Avenir" (EUR INTREE, reference ANR-18-EURE-0010) and PRC ANR  MagDesign and TcPredictor. The computing resources are provided by ACCESS (TG-DMR180040) and High-Performance Computing Centre Adastra/CINES of GENCI (projects A0140807539 and A0160815101). We also acknowledge the reviewers for their insightful comments and suggestions during the review stage.
During the preparation of this work, the authors used GitHub Copilot in order to improve the code readability and documentation. After using this tool/service, the authors reviewed and edited the content as needed and take full responsibility for the content of the publication.

\section*{Data availability}
The \texttt{LEGO\_xtal} source code, instructions, as well as scripts used to calculate the results of this study, are available in \url{https://github.com/MaterSim/LEGO-xtal}. The complete list of 1741 low-energy sp$^2$ carbon allotropes, as well as ReaxFF, MACE and DFT-r$^2$SCAN energies, is listed in \url{http://lego-crystal.onrender.com}.

\section*{Conflict of interest}
All authors declare that they have no conflict of interest.

\section*{Author contributions}
The concept of \textit{Crystal Structure Generator From Chemical Building Blocks} was initially discussed by G.F. and Q.Z. (with the latter being a Visiting Professor at Poitiers University in June 2023). Q.Z. and G.F. co-conceived the idea. With the help of G.F, Q.Z. initiated the framework. Q.Z., D.D., H.X., and G.F. supervised this project. O.G.R., M.S.R., H.X., and Q.Z. implemented the code, S.P. performed the electronic structure calculations. All authors analyzed the results and contributed to manuscript writing.

\clearpage
\setcounter{section}{0}
\setcounter{figure}{0}
\setcounter{table}{0}
\setcounter{equation}{0}
\setcounter{lstlisting}{0}
\renewcommand{\thesection}{S\arabic{section}}
\renewcommand{\thefigure}{S\arabic{figure}}
\renewcommand{\thetable}{S\arabic{table}}
\renewcommand{\theequation}{S\arabic{equation}}
\renewcommand{\thelstlisting}{S\arabic{lstlisting}}
\begin{center}
\textbf{\Large Supplementary Materials}
\end{center}

\section{Generative Models Training and Sampling}
\subsection{Generative Adversarial Network (GAN)}
Generative Adversarial Networks (GANs) have emerged as powerful generative models capable of synthesizing realistic data by training two neural networks—the Generator and the Discriminator—in an adversarial setting \cite{goodfellow2014generative}. The Generator aims to create synthetic data that closely resembles real data, while the Discriminator is responsible for distinguishing between real and synthetic inputs. During training, both components are optimized in an adversarial process: the Generator progressively improves its ability to produce data that mimics real samples with increasing fidelity, while the Discriminator becomes more adept at detecting even subtle discrepancies between real and generated data. This dynamic continues until an equilibrium is reached, at which point the Generator produces outputs so realistic that the Discriminator can no longer reliably differentiate them from real data. The GAN model in our \texttt{LEGO-xtal} framework is directly inspired by the Conditional Tabular GAN (CTGAN) approach proposed by Xu et al. \cite{ctgan}, specifically designed to handle structured tabular data containing both discrete and continuous attributes. 

\paragraph{Generator.}
The Generator is responsible for transforming random noise into realistic tabular data samples that resemble the true distribution. This noise, denoted as \( z \sim \mathcal{N}(0, I) \), is drawn from a standard multivariate normal distribution. The Generator architecture is composed of multiple residual blocks. Each block includes a fully connected (FC) linear transformation, followed by batch normalization (BN) \cite{ioffe2015batch}, and a Rectified Linear Unit (ReLU) activation function. Residual (or skip) connections \cite{he2016deep} are used to concatenate the input of each block with its output, helping preserve information flow and mitigate vanishing gradients during training.
The internal computations of the Generator can be summarized as follows:
\begin{equation}
\begin{cases}
h_0 &= \mathbf{z} \in \mathbb{R}^{B \times 128} \\[4pt]
r_1 &= \mathrm{ReLU} \circ \mathrm{BN} \circ \mathrm{FC}_{128 \rightarrow 512}(h_0) \\[4pt]
h_1 &= r_1 \oplus h_0 \in \mathbb{R}^{B \times 640} \\[6pt]
r_2 &= \mathrm{ReLU} \circ \mathrm{BN} \circ \mathrm{FC}_{640 \rightarrow 512}(h_1) \\[4pt]
h_2 &= r_2 \oplus h_1 \in \mathbb{R}^{B \times 1152} \\[6pt]
\hat{\mathbf{x}} &= \mathrm{FC}_{1152 \rightarrow d_x}(h_2)
\end{cases}
\end{equation}

\noindent
Here,
\begin{itemize}
    \item \( \mathbf{z} \): Latent noise vector sampled from a standard normal distribution.
    \item \( B \): Batch size.
    \item \( \mathrm{FC}_{a \rightarrow b} \): Fully connected (linear) layer with input dimension \( a \) and output dimension \( b \).
    \item \( \mathrm{BN} \): Batch normalization layer.
    \item \( \mathrm{ReLU} \): Rectified Linear Unit activation function.
    \item {\( \oplus \): Concatenation operator along the feature dimension.}
    \item \( \circ \) : Element-wise multiplication
    \item \( \hat{\mathbf{x}} \): Output vector with dimensionality matching the trainee input data (\( d_x \)).
\end{itemize}
The generated samples lie in a transformed feature space comprising both continuous and categorical components. To handle these different types appropriately, we apply distinct activation functions based on metadata recorded during preprocessing. Each feature is tagged with a label, either \texttt{tanh} or \texttt{gumbel}, indicating the appropriate postprocessing strategy.

For continuous features, we use \textit{cluster-based normalization}, implemented via Gaussian Mixture Modeling (GMM) \cite{patki2016synthetic}. Each feature is modeled as a mixture of Gaussians, and transformed into a combination of two components: (1) a scalar normalized value indicating the position within a selected Gaussian, and (2) a one-hot encoded vector representing the most likely cluster. During generation, the scalar component is passed through the hyperbolic tangent activation function, \( \tanh \), to restrict its values to the range \((-1, 1)\), maintaining numerical stability and aligning with the transformed scale.

For categorical (discrete) features, which are represented using one-hot encoding, the Gumbel-Softmax function \cite{jang2017categorical} is applied. This enables differentiable sampling from a categorical distribution, allowing training via backpropagation despite the discrete nature of these variables. The Gumbel-Softmax activation ensures the synthetic outputs closely follow the one-hot format of the original data.

Finally, a composite loss is computed by comparing the activated outputs to the real transformed inputs. Separate loss components are used for continuous and categorical features, enabling the model to learn an accurate and data-type-aware generative process.

\paragraph{Discriminator.}
The Discriminator is a neural network trained to distinguish between real data and data generated by the Generator. It plays a crucial role in guiding the Generator to produce more realistic samples. In our architecture, the Discriminator consists of fully connected (dense) layers with LeakyReLU activations, which help stabilize learning and reduce the chance of neurons becoming inactive during training \cite{maas2013rectifier}. Additionally, Dropout layers are included to prevent overfitting, improving the model’s ability to generalize to unseen samples.

\noindent
We also adopt the \textit{packing} strategy (\texttt{PAC}) \cite{ctgan} which improves training stability. Instead of feeding single samples into the Discriminator, we group every 10 samples together and concatenate them into one larger vector. This allows the Discriminator to evaluate patterns across sample groups rather than treating them independently.

\noindent
Mathematically, the Discriminator operates as follows:

\[
\begin{cases}
x_{\text{pac}} = x \in \mathbb{R}^{B \cdot 10 \times d_x} \rightarrow \text{reshape} \in \mathbb{R}^{B \times (10 \cdot d_x)} \\[5pt]
h_1 = \text{Dropout} \circ \text{LeakyReLU}_{\gamma=0.2} \left( \text{FC}_{10 \cdot d_x \to 512}(x_{\text{pac}}) \right) \\[5pt]
h_2 = \text{Dropout} \circ \text{LeakyReLU}_{\gamma=0.2} \left( \text{FC}_{512 \to 512}(h_1) \right) \\[5pt]
y = \text{FC}_{512 \to 1}(h_2)
\end{cases}
\]

\noindent
Here, \( x_{\text{pac}} \) denotes the packed input with batch size \( B \), and each group of 10 samples of original feature dimension \( d_x \) is reshaped into a vector of size \( 10 \cdot d_x \). This reshaped input then flows through two fully connected layers with Dropout (0.2) and LeakyReLU activation functions.

\noindent
Using a small slope for negative values helps avoid the ``dying ReLU'' \cite{maas2013rectifier} problem by allowing gradients to pass through even when the input is negative, ensuring the network continues to learn effectively.
\paragraph{Loss Functions.}

Our model utilizes the Wasserstein GAN with Gradient Penalty (WGAN-GP) loss formulation for the Discriminator \cite{gulrajani2017improved}. Specifically, the Discriminator loss \( L_D \) is defined as:
\begin{equation}
    L_{D} = [D(\tilde{x})] - [D(x)] + \lambda \left[(\|\nabla_{\hat{x}}D(\hat{x})\|_2 - 1)^2\right]
\end{equation}
where \( x \) represents real samples, \( \tilde{x} \) represents generated samples, \( \hat{x} \) are interpolated samples between real and generated samples, \( D \) is the discriminator function, \( \nabla_{\hat{x}}D(\hat{x}) \) denotes the gradient with respect to \( \hat{x} \), and \( \lambda \) is the gradient penalty coefficient.
The Generator loss \( L_G \) aims to minimize the discriminator’s ability to distinguish generated samples as fake and is defined as: 

\begin{equation}
    L_{G} = -[D(\tilde{x})]
\end{equation}

\begin{figure}[h]
    \centering
    \includegraphics[width=0.5\textwidth]{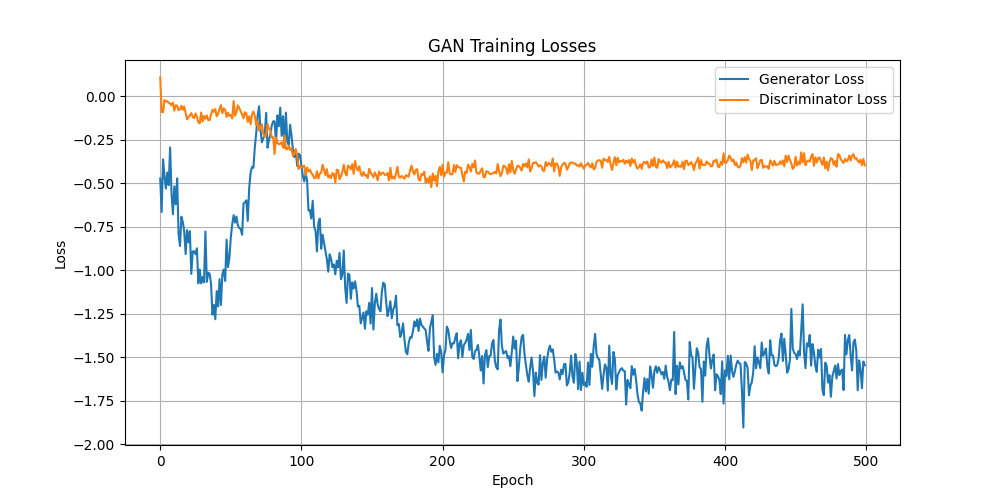}
    \caption{\textbf{GAN training losses over 500 epochs.} The generator loss (blue) improves over time, while the discriminator loss (orange) stabilizes.}
    \label{fig:gan_training_loss}
\end{figure}

\paragraph{Training Model.}
To enable reproducibility of our results, we provide a simple Python script demonstrating how to retrain the GAN-based synthesizer on the processed training data. The model is implemented using our \texttt{lego.GAN} module and supports discrete column conditioning. Below is an example of how the model can be trained and used to generate new samples:

\begin{lstlisting}[language=Python, caption=A Python script to train the GAN model in the \texttt{LEGO-xtal} Framework., label=lst:subgroup-opt]
# https://github.com/MaterSim/LEGO-xtal
from lego.GAN import GAN
import pandas as pd

# Load processed training data
df = pd.read_csv('data/train/train-v4.csv')

# Specify discrete columns used for conditioning
discrete_columns = ['spg','wp0','wp1','wp2','wp3','wp4','wp5','wp6','wp7']

# Initialize and train the synthesizer
synthesizer = GAN()
synthesizer.fit(df, discrete_columns=discrete_columns)

# Sample synthetic data and save to file
df_synthetic = synthesizer.sample(samples=1000)
df_synthetic.to_csv('synthetic_structures_GAN.csv', index=False, header=True)
\end{lstlisting}
During training, the generator and discriminator losses were recorded across 500 epochs. The plot in Figure~\ref{fig:gan_training_loss} illustrates the convergence behavior of the GAN. The generator loss steadily improves, while the discriminator loss stabilizes, indicating a balanced adversarial training process.

\subsection{Variational Autoencoder (VAE)}
The Variational Autoencoder (VAE) is a deep generative model that learns a probabilistic mapping between high-dimensional observed data and a lower-dimensional latent space from which synthetic data can be generated \cite{kingma2013auto, rezende2014stochastic}. The VAE consists of two neural networks: an encoder that transforms input data into the parameters of a latent Gaussian distribution, and a decoder that reconstructs data from samples drawn from this distribution. To enable gradient-based optimization, the reparameterization trick is employed, which makes the sampling process differentiable.

In this work, we utilize the Tabular Variational Autoencoder (TVAE) architecture as implemented in the open-source SDV project \cite{patki2016synthetic}, which extends the VAE framework to structured tabular datasets. We adopt this architecture within our proposed \texttt{LEGO-xtal} framework to model the underlying distribution of crystallographic symmetry features and generate novel candidate crystal structures.

\paragraph{Encoder \(\,q_{\phi}(z \mid x)\).}
Let \( \mathbf{x} \in \mathbb{R}^{B \times d_x} \) denote a batch of input data with \(B\) samples, each having \(d_x\) features. In the encoder network, the input data \( \mathbf{x} \) is passed through two FC + ReLU layers to produce a hidden representation \( h_2 \). From this, two separate FC layers compute the mean \( \mu \in \mathbb{R}^{B \times 128} \) and the log-variance \( \log \boldsymbol{\sigma}^2 \in \mathbb{R}^{B \times 128} \) of the latent distribution.

\begin{equation}
\begin{cases}
x &= \mathbf{x} \in \mathbb{R}^{B \times d_x} \\[4pt]
h_1 &= \mathrm{ReLU} \circ \mathrm{FC}_{d_x \rightarrow 512}(x) \\[2pt]
h_2 &= \mathrm{ReLU} \circ \mathrm{FC}_{512 \rightarrow 512}(h_1) \\[4pt]
\mu &= \mathrm{FC}_{512 \rightarrow 128}(h_2) \\[2pt]
\log \boldsymbol{\sigma}^2 &= \mathrm{FC}_{512 \rightarrow 128}(h_2) \\[4pt]
\end{cases}
\end{equation}

 To sample the latent vector \( \mathbf{z} \), the reparameterization trick is used:
\[
\mathbf{z} = \mu + \exp\left( \tfrac{1}{2} \log \boldsymbol{\sigma}^2 \right) \odot \boldsymbol{\epsilon}, \quad \boldsymbol{\epsilon} \sim \mathcal{N}(0, \mathbf{I})
\]
where \( \odot \) denotes element-wise multiplication.
\paragraph{Decoder \(\,p_{\theta}(x \mid z)\).}
The decoder takes the latent variable \( \mathbf{z} \in \mathbb{R}^{B \times 128} \) and reconstructs the input using two additional FC + ReLU layers, followed by a final FC layer producing the output \( \hat{\mathbf{x}}_{\text{out}} \in \mathbb{R}^{B \times d_x} \). Additionally, a learnable parameter vector \( \boldsymbol{\sigma}_{\text{out}} \in [0.01, 1.0]^{d_x} \) is used to model the per-feature uncertainty for continuous-valued outputs. This parameter is jointly optimized during training to improve reconstruction quality.

\begin{equation}
\begin{cases}
z &= \mathbf{z} \in \mathbb{R}^{B \times 128} \\[4pt]
h_1 &= \mathrm{ReLU} \circ \mathrm{FC}_{128 \rightarrow 512}(z) \\[2pt]
h_2 &= \mathrm{ReLU} \circ \mathrm{FC}_{512 \rightarrow 512}(h_1) \\[4pt]
\hat{\mathbf{x}}_{\text{out}} &= \mathrm{FC}_{512 \rightarrow d_x}(h_2) \\[6pt]
\end{cases}
\end{equation}

\paragraph{Loss Function.}
The training objective of the VAE model combines two components: the reconstruction loss and the Kullback-Leibler divergence (KLD). Given an input batch \( \mathbf{x} \in \mathbb{R}^{B \times d_x} \), the VAE aims to reconstruct \( \mathbf{x} \) from the latent space while ensuring that the learned posterior distribution remains close to a unit Gaussian prior.

For each input feature, the type of reconstruction loss is chosen based on the activation function defined in the data transformer. The total loss is computed as:

\begin{equation}
\mathcal{L}_{\text{total}} = \frac{\lambda}{B} \sum_{i=1}^B \mathcal{L}_{\text{recon}}^{(i)} + \frac{1}{B} \sum_{i=1}^B D_{\text{KL}} \left( q_\phi(z^{(i)} \mid x^{(i)}) \,\|\, \mathcal{N}(0, \mathbf{I}) \right)
\end{equation}

\subparagraph{Reconstruction Loss \(\mathcal{L}_{\text{recon}}\).}
The reconstruction loss is computed feature-wise:

- For continuous features, we model the output as a Gaussian distribution with mean predicted by the decoder and learnable standard deviation \(\sigma_j \in [0.01, 1.0]\), producing:

\begin{equation}
\mathcal{L}_{\text{cont}} = \sum_{j \in \mathcal{C}} \left( \frac{(x_{\text{in}} - \tanh(\hat{x}_{\text{out}}))^2}{2 \sigma_{\text{out}}^2} + \log \sigma_{\text{out}} \right)
\end{equation}

Here, \(\mathcal{C}\) denotes the set of continuous feature indices, and \(\hat{x}_{\text{out}}\) is the raw output of the decoder before activation. The \(\tanh\) function ensures bounded output for numeric stability.

- For categorical features, the cross-entropy loss is applied between the target one-hot vector \(x_{\text{out}}\) and the predicted logits \(\hat{x}_{\text{in}}\):

\begin{equation}
    \mathcal{L}_{\text{cat}} = \sum_{j \in \mathcal{D}} \text{CrossEntropy}\left( \hat{x}_{\text{out}}, \arg\max(x_{\text{in}}) \right)
\end{equation}

Here, \(\mathcal{D}\) denotes the set of categorical feature indices.

\subparagraph{KL Divergence.}
The Kullback-Leibler divergence encourages the encoder's learned latent distribution \( q_\phi(z \mid x) = \mathcal{N}(\mu, \operatorname{diag}(\sigma^2)) \) to remain close to the prior \( \mathcal{N}(0, \mathbf{I}) \):

\begin{equation}
    D_{\text{KL}} = -\frac{1}{2} \sum_{k=1}^{d_z} \left(1 + \log \sigma_k^2 - \mu_k^2 - \sigma_k^2\right)
\end{equation}

\noindent
The total loss is scaled by a user-defined factor \(\lambda\) (default: 2.0) to balance reconstruction and regularization effects.

\begin{figure}[h]
    \centering
    \includegraphics[width=0.5\textwidth]{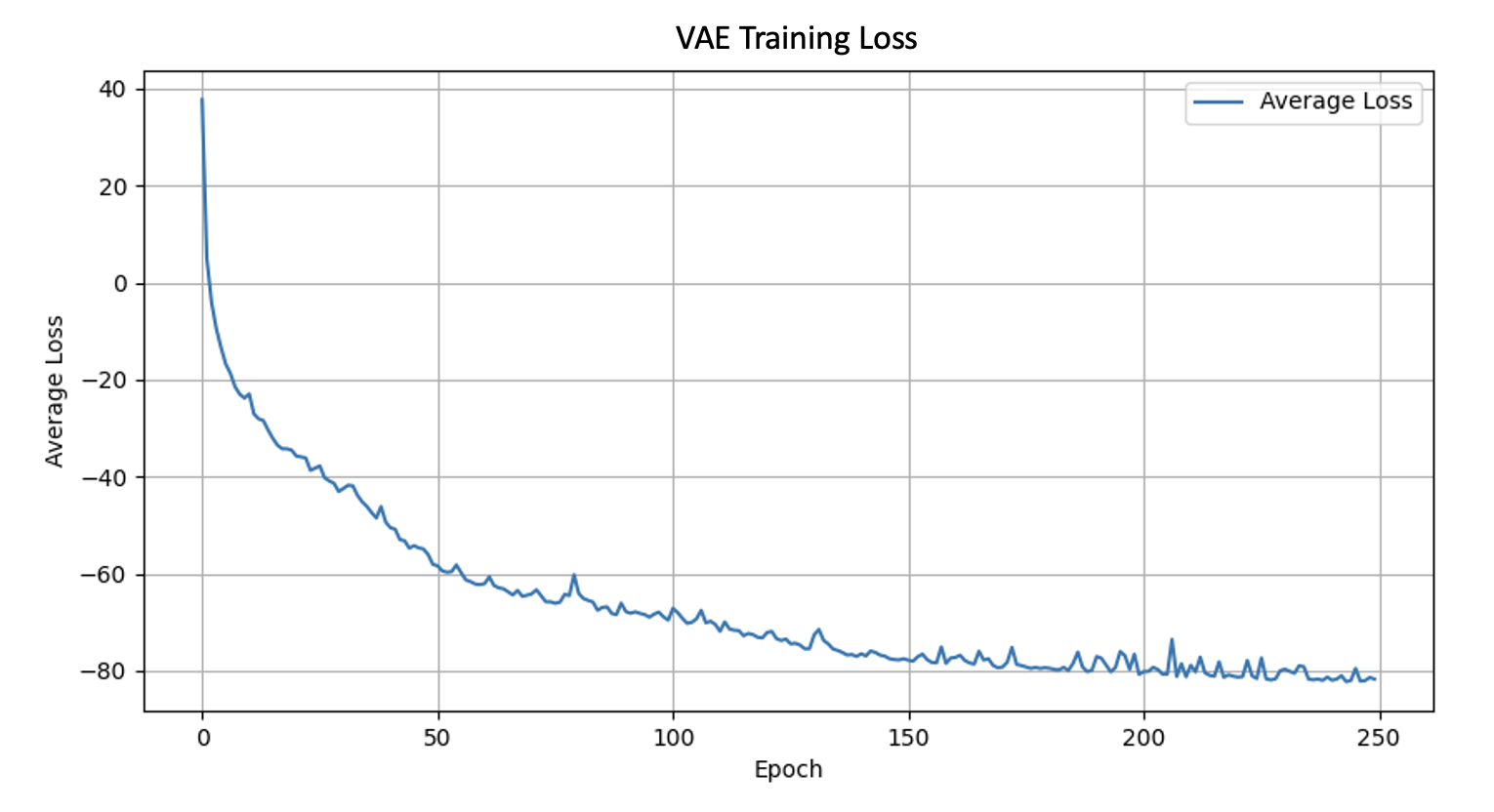}
    \caption{\textbf{The VAE Model training loss over 250 epochs.}}
    \label{fig:vae_loss}
\end{figure}

\paragraph{Training.}
The model is trained for 250 epochs using the Adam optimizer~\cite{kingma2017adammethodstochasticoptimization} with L2 weight regularization. Input data is preprocessed, which contains normalized continuous features and one-hot encoded categorical features. During training, the loss is computed on mini-batches of size 500. To ensure numerical stability, the per-feature variance parameters \( \boldsymbol{\sigma}_{\text{out}} \) are clamped between 0.01 and 1.0. The average training loss per epoch is recorded, showing a smooth convergence trend (see Figure~\ref{fig:vae_loss}). Below is an example of how the model can be trained
and used to generate new samples using the \texttt{LEGO-xtal} framework:

\begin{lstlisting}[language=Python, caption=A Python script to train the VAE model in the \texttt{LEGO-xtal} Framework., label=lst:subgroup-opt]
# https://github.com/MaterSim/LEGO-xtal
from lego.VAE import VAE
import pandas as pd

# Load processed training data
df = pd.read_csv('data/train/train-v4.csv')

# Specify discrete columns used for conditioning
discrete_columns = ['spg','wp0','wp1','wp2','wp3','wp4','wp5','wp6','wp7']

# Initialize and train the synthesizer
synthesizer = VAE()
synthesizer.fit(df, discrete_columns=discrete_columns)

# Sample synthetic data and save to file
df_synthetic = synthesizer.sample(samples=1000)
df_synthetic.to_csv('synthetic_structures_VAE.csv', index=False, header=True)
\end{lstlisting}

\section{The SO(3) Descriptor} 

In 2012, Bartók et al. introduced an improved many-body descriptor that explicitly incorporates both radial and angular components \cite{Bartok-PRB-2013}. This approach overcomes the limitation of describing neighbor density with a Dirac delta function by replacing the delta function with a Gaussian function of limited width $\alpha$. This smoothing allows for a more realistic representation of how atoms are distributed around a reference atom. According to Bartók, the modified neighbor density function is given by:

\begin{equation}
\rho^\prime(\mathbf{r}) = \sum_i^{r_i \leq r_c} e^{(-\alpha|\mathbf{r}-\mathbf{r_i}|^2)}
= \sum_i^{r_i \leq r_c} e^{-\alpha(r^2+r_i^2)} e^{2\alpha \mathbf{r} \cdot \mathbf{r_i}} 
\end{equation}

Expanding the exponential of a dot product in spherical coordinates:

\begin{equation}
e^{2\alpha \mathbf{r} \cdot \mathbf{r_i}} = e^{2\alpha r r_i \mathrm{cos(\gamma)}} = 4\pi \sum_{l=0}^{\infty} \sum_{m=-l}^{l} I_l(2\alpha r r_i) Y_{lm}^{*}(\hat{\mathbf{r_i}}) Y_{lm}(\hat{\mathbf{r}}).    
\end{equation}

In which, we used the general formula addition theorem for spherical harmonics,

\begin{equation}
e^{z \mathrm{cos(\gamma)}} = 4\pi \sum_{l=0}^{\infty} \sum_{m=-l}^{l} I_l(z) Y_{lm}^{*}(\hat{\mathbf{r_i}}) Y_{lm}(\hat{\mathbf{r}}).    
\end{equation}

This expression can be further expanded as:

\begin{equation}
\rho^\prime(\mathbf{r}) = \sum_i^{r_i \leq r_c} \sum_{lm}  4\pi e^{-\alpha(r^2+r_i^2)} I_l(2\alpha r r_i) Y_{lm}^*(\mathbf{\hat{r_i}}) Y_{lm}(\mathbf{\hat{r_i}}),    
\end{equation}

where the first part $I_l(2\alpha r r_i)$ is the modified spherical Bessel function of the first kind (governed by $2\alpha r r_i$), providing the radial dependence, and the second part captures the angular dependence of the vectors $\mathbf{r}$ and $\mathbf{r}_i$.

Bartók also introduced a set of polynomials, $g_n(r)$, which help describe the radial component in a more refined way:

$$
\phi_\alpha(r) = (r_c - r)^{\alpha +2}/N_\alpha
$$

where $N_\alpha$  is a normalization factor given by:
\[
N_\alpha = \sqrt{\int_0^{r_c} r^2(r_c-r)^{2(\alpha+2)}dr}
\]

These polynomials are orthonormalized to ensure that the radial functions $g_n(r)$ form a basis. The orthonormalization process is performed through linear combinations of $\phi_\alpha(r)$, and the coefficients are obtained from 

\[
g_n(r) = \sum_{\alpha=1}^{n_{\textrm{max}}}W_{n\alpha}\phi_\alpha(r),
\]

where $W$ is constructed from the inverse square root of the overlap matrix $S$,

\begin{align*}
    S_{\alpha\beta} 
    &= \int_0^{r_c}r^2\phi_\alpha(r)\phi_\beta(r)dr \\ 
    &= \frac{\sqrt{(2\alpha+5)(2\alpha+6)(2\alpha+7)(2\beta+5)(2\beta+6)(2\beta+7)}}{(5+\alpha+\beta)(6+\alpha+\beta)(7+\alpha+\beta)}    
\end{align*}

This overlap matrix describes how different radial functions overlap with each other and ensures that the final radial basis functions $g_n(r)$ are orthonormal.

The neighbor density function $\rho^{\prime}(\mathbf{r})$ can then be expanded in terms of both the radial basis $g_n(r)$ and the spherical harmonics:

\begin{align}
c_{nlm} &= \left< g_n(r)Y_{lm}(\mathbf{\hat{r}})|\rho^{\prime}(\mathbf{r}) \right> \\\nonumber
&= \int d^3r g_n(r) Y_{lm}(\hat{\mathbf{r}}) ) 
\sum_{r_i \leq r_c} \sum_{l^{\prime}m^{\prime}} 4\pi e^{-\alpha(r^2 + r_i^2)} I_{l^{\prime}}(2\alpha r r_i) Y_{l^{\prime}m^{\prime}}^{*}(\hat{\mathbf{r_i}}) Y_{l^{\prime}m^{\prime}}(\hat{\mathbf{r}})
\end{align}

When integrating over the angular variables $\hat{\mathbf{r}}$, only the terms with $l^{\prime} = l$ and $m^{\prime} = m$  will survive, due to orthogonality. 

\begin{align}
c_{nlm} &= 4\pi \sum_i^{r_i \leq r_c} Y_{lm}^* (\hat{\mathbf{r_i}})  \int_0^{r_c} r^2 g_n(r) e^{-\alpha(r^2 + r_i^2)} I_l(2\alpha r r_i) dr\\\nonumber
&= 4\pi \sum_i^{r_i \leq r_c} e^{-\alpha r_i^2} Y_{lm}^*(\mathbf{\hat{r}_i}) 
\int_0^{r_c} r^2 g_n(r) e^{-\alpha r^2} I_l(2\alpha r r_i) dr    
\end{align}

Finally, the rotation-invariant power spectrum is obtained by combining these expansion coefficients:

\begin{equation}
p_{n_1 n_2 l} = \sum_{m=-l}^{+l} c_{n_1 l m} c^*_{n_2 l m},
\end{equation}

where the expansion coefficients $c_{nlm}$ are projections of the neighbor density onto a set of orthonormal radial basis functions $g_n(r)$ and spherical harmonics $Y_{lm}(\hat{\mathbf{r}})$. This descriptor effectively encodes the local atomic geometry, including coordination shells and angular distributions, while ensuring invariance to global rotations.

It has been suggested that pure spherical harmonics descriptor cannot reconstruct the frequency components uniquely up to rotation \cite{kazhdan2003rotation}.
To remedy this issue, the SO(3) descriptor can be further complemented with the radial distribution function (RDF) to capture the radial distribution of atoms around a central atom. The RDF is defined as:
\begin{equation}
g(r) = \frac{1}{N} \sum_{i=1}^{N} \sum_{j \neq i} \delta(r - |\mathbf{r}_i - \mathbf{r}_j|),
\end{equation}
where $N$ is the total number of atoms, $\mathbf{r}_i$ and $\mathbf{r}_j$ are the positions of atoms $i$ and $j$, and $\delta$ is the Dirac delta function. The RDF provides additional information about the distribution of atoms at various distances from a central atom, complementing the SO(3) descriptors angular information.

\section{Pre-relaxation}

\subsection{Descriptor-based Geometry Optimization}
After generating a crystal structure with the desired local environment, the next step is to optimize the geometry of the crystal structure. The optimization process is crucial for refining the generated structure to ensure it is energetically favorable and stable. In this work, we employ a descriptor-based optimization approach that utilizes the SO(3) descriptor to guide the optimization process via \texttt{scipy.optimize.minimize}.

\begin{itemize}
    \item \textbf{Nelder--Mead} (100 iterations): a gradient-free simplex algorithm that is robust for local non-smooth landscapes and suitable for early-stage refinement,
    \item \textbf{L-BFGS-B} (100 iterations): a quasi-Newton method with bound constraints, used for smooth, efficient convergence in later stages.
    \item \textbf{Basin-hopping}: optionally used for global optimization, combining local minimization with random jumps to escape local minima.
\end{itemize}
More details about the optimization can be found in the documentation of \textit{pyxtal.lego.builder.optimize\_xtal} \cite{pyxtal}.

\subsection{Geometry Optimization under Subgroup Representation}

\begin{figure*}[htbp]
    \centering
    \includegraphics[width=0.9\textwidth]{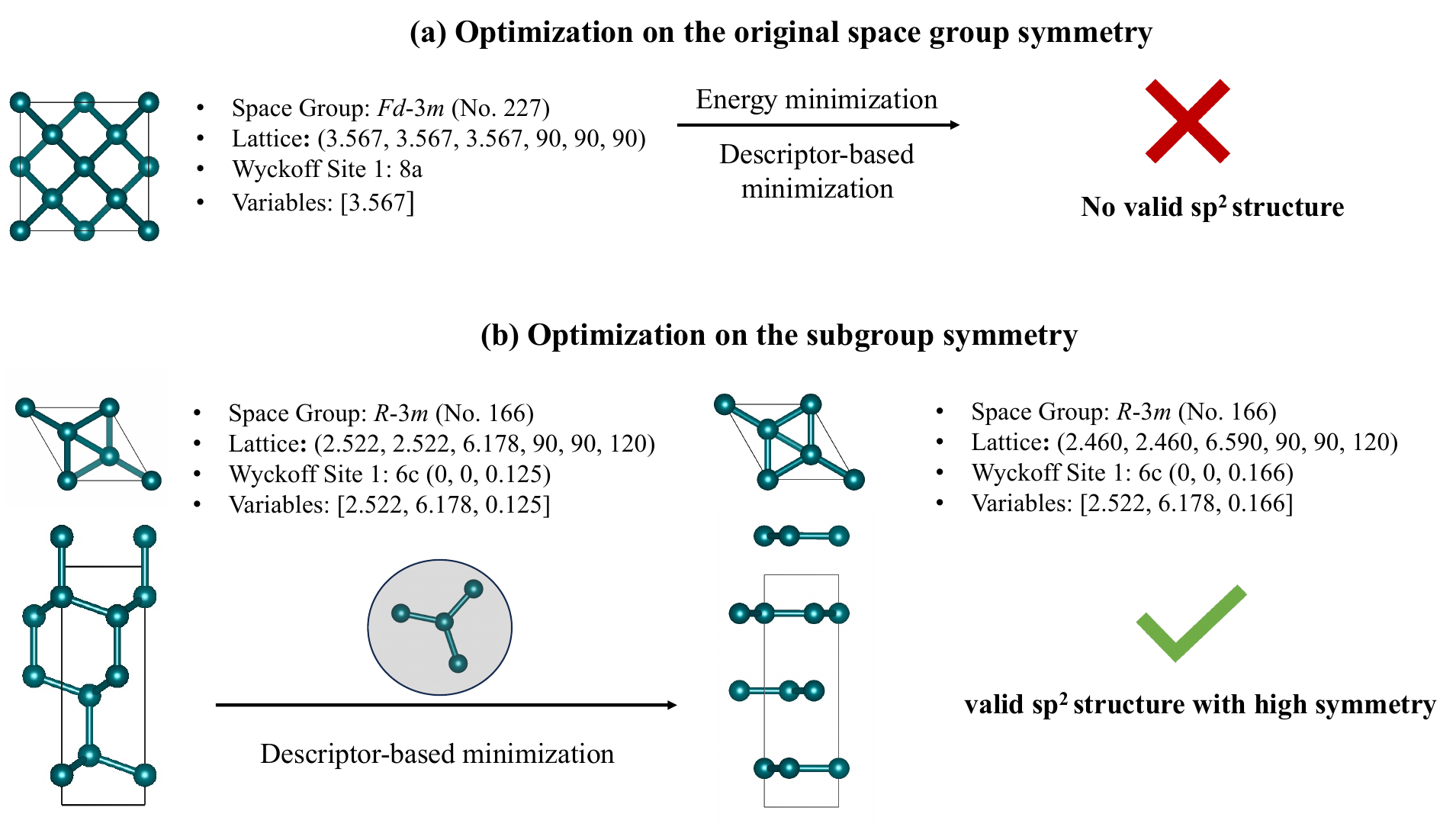}
    \caption{\textbf{Geometry optimization under the symmetry constraint from diamond structure to the sp$^2$ environment.} (a) optimization from the original $Fd$-3$m$ symmetry; (b) optimization from the subgroup $R$-3$m$ symmetry.}
    \label{fig:sub-opt}
\end{figure*}

Figure \ref{fig:sub-opt} illustrates how to optimize the geometry of a crystal structure under subgroup representation. If one perform the symmetry-constrained optimization starts from a diamond structure with $Fd$-3$m$ symmetry, neither the local environment nor the energy landscape is suitable to obtain the sp$^2$ allotropes. If one starts from a subgroup $R$-3$m$ symmetry, the energy-based optimization will retain the diamond structure packing with the sp$^3$ local environment. However, the descriptor-based optimization can lead to a local environment that is more compatible with sp$^2$ allotropes. This is because the subgroup representation allows for a more reduced crystal variables that allows for the atomic transition to the sp$^2$ allotropes, while the original $Fd$-3$m$ symmetry is too rigid and does not allow for such flexibility. Using the subgroup representation, one can effectively tune the local environment even for a low-energy crystal structure with unwanted local environment.

The above demonstration can be achieved using the following codes (Listing \ref{lst:subgroup-opt}) based on \texttt{PyXtal}.

\begin{lstlisting}[language=Python, caption=A Python script to perform geometry optimization using the subgroup representation., label=lst:subgroup-opt]
# pip install pyxtal
from pyxtal.lego.builder import builder
from pyxtal import pyxtal

# Get the graphite reference environment and set up optimizer
xtal = pyxtal()
xtal.from_prototype('graphite')
cif_file = xtal.to_pymatgen()
bu = builder(['C'], [1], db_file='test.db')
bu.set_descriptor_calculator(mykwargs={'rcut': 2.0})
bu.set_reference_enviroments(cif_file)
print(bu)

# Get the diamond crystal
xtal = pyxtal()
xtal.from_prototype('diamond')
print(xtal)
print(xtal.get_1d_rep_x())
_, _, _ = bu.optimize_xtal(xtal)

# Subgroup conversion
sub_xtal = xtal.subgroup_once(H=166, eps=1e-4)
sub_xtal.to_file('sp3.cif')
print(sub_xtal)
print(sub_xtal.get_1d_rep_x())

# Optimization
sub_xtal, loss, _ = bu.optimize_xtal(sub_xtal)
print(sub_xtal)
print(sub_xtal.get_1d_rep_x())
sub_xtal.to_file('sp2.cif')
\end{lstlisting}

\subsection{Pre-relaxation Bias Analysis}
As we described in the main text, we
propose the use of pre-relaxation to achieve two goals
\begin{enumerate}
    \item  Maintain the same chance if the initial guess is close to an ideal arrangement,
    \item Enhance the chance of generating good structures even from a bad guess by pre-relaxation.
\end{enumerate}

To verify if the pre-relaxation process introduces unwanted bias, we first performed a local perturbation (up to 0.5 \AA) to the 140 known sp$^2$ carbon allotropes. After the pre-relaxation, all of structures returned to the original geometry, suggesting that this approach yields similar relaxation outcome to the energy-based approach when the initial structure is close to the ideal sp$^2$ environment. When the initial guess is far from the ideal case, the pre-relaxation not only saves a lot of computational time, but also effectively further reduces the searching space by excluding the possibility of handling low-energy configurations with an undesired local environment.

Furthermore, Fig. \ref{fig:aug} and Fig. \ref{fig:overview} of the main text show two pre-relaxation examples that turn the seemingly irrelevant structures to perfect sp$^2$-bonding carbon networks. This may look striking from the energy perspective since the initial chemical bonding is hard to break if one seeks to minimize the energy. However, it is feasible if the target is to optimize the geometry as defined by the descriptor. In Fig. \ref{fig:sub-opt}, we also show that one can even relax a perfect sp$^3$-based diamond structure to a sp$^2$ network.

In summary, we indeed introduced biases to force the system to transform to the desired local environment. However, this should not create artifacts if the initial guess is close to the target environment.

\subsection{Comparison of Different Relaxation Strategies}
In Table \ref{tab:time}\ref{tab:time} of the manuscript, we report the timing for the MACE or ReaxFF based on the remaining valid sp2 structures after pre-relaxation. The use of pre-relaxation is crucial for the success of the subsequent MACE or ReaxFF relaxation. This is because the VAE/GAN generated structures may contain unphysical features, such as very short interatomic distances or highly distorted bond angles, which can lead to convergence issues or unrealistic results during direct relaxation with MACE or ReaxFF. The pre-relaxation step helps to correct these issues by adjusting atomic positions and lattice parameters to more physically reasonable values, thereby providing a better starting point for the more accurate but computationally intensive MACE or ReaxFF methods.
If we directly use MACE or ReaxFF to perform the relaxation on the VAE/GAN generated samples, both the timing and success rate will become significantly worse.

To illustrate this concept, we used the VAE-cont-60k model to generate 10K structures, and then relaxed them with (1) descriptor-based pre-relaxation followed by GULP relaxation; (2) descriptor-based pre-relaxation followed by MACE relaxation; (3) GULP-ReaxFF only relaxation and (4) MACE relaxation only. The results are summarized in Table \ref{tab:relaxation_comparison}.

\begin{table}[h]
    \centering
    \caption{Comparison of relaxation strategies for 1k samples on a 96-core CPU node.}
    \begin{tabular}{|l|c|c|}
        \hline
        \textbf{Method} & \textbf{Time per 1k samples} & \textbf{Valid unique sp$^2$} \\
        \textbf{Method} & \textbf{(mins)} & \textbf{structures (\%)} \\
        \hline
        Pre-relax + GULP & 5.5 + 1.5 & 7.83 \\
        Pre-relax + MACE & 5.5 + 2.5 & 7.83 \\
        GULP-only        & $>$20 & 1.03 \\
        MACE-only        & $>$120 & 1.83 \\
        \hline
    \end{tabular}
    \label{tab:relaxation_comparison}
\end{table}

Clearly, the GULP-only and MACE-only relaxation strategies are approximately 3 and 15 times slower, respectively, than those incorporating pre-relaxation. For both GULP-only and MACE-only relaxations, a timeout of 60 minutes was enforced to prevent calculations from stalling—a common occurrence for random structures. Furthermore, if a significant fraction of very complex structures (with more than 100 atoms per unit cell) is present, the actual computational cost may increase to 5-10 for GULP and 15–50 for MACE.

More importantly, direct relaxation using MACE or GULP-ReaxFF often yields low-energy structures with mixed sp, sp$^2$, and sp$^3$ environments, resulting in a low success rate (less than 2\%) for generating valid sp$^2$ structures, as compared to the 7.83\% with the inclusion of pre-relaxation. This success rate is expected to decrease further when the initial structure contains multiple Wyckoff sites. Thus, the pre-relaxation step not only accelerates the process but also substantially improves the success rate.

Last, the batch-wise relaxation has been implemented, reducing the pre-relaxation time from 5 minutes per 1,000 samples to less than 1–2 minutes per 1,000 samples under the GPU environment. These improvements are available at \href{https://github.com/MaterSim/LEGO-xtal/tree/main/lego/Batch_Pre-relaxation}{LEGO-xtal} repository and will continue to be refined. Conceptually, further speed enhancements are anticipated as GPU acceleration is adopted.

\subsection{Pre-relaxation Examples on the Multi-component SiO\textsubscript{2} System}

While the descriptor-based pre-relaxation method has been primarily demonstrated on single-element systems, it is equally applicable to multi--component systems. The key requirement is the availability of a reference structure that accurately represents the local environments of all constituent elements. This reference structure serves as a template for defining the desired coordination environments and bonding characteristics for each element in the multi--component system.

To demonstrate the pre-relaxation process for multi--component systems, we provide an example using SiO\textsubscript{2}. The reference structure is set to $\alpha$-cristobalite, which contains two Wyckoff positions: one for Si (4a) and another for O (8b). The coordination numbers are specified as 4 for Si and 2 for O, with the exclusion of direct O-O bonding. The builder is configured to generate structures with space group 154 ($P3_121$) and Wyckoff positions 3a and 6c. The optimization process is performed using the descriptor-based method. The complete code is provided in Listing \ref{lst:sio2}.

\begin{lstlisting}[language=Python, caption=A Python script to handle the SiO$_2$ system, label=lst:sio2]
# pip install pyxtal
from pyxtal.lego.builder import builder
from pyxtal import pyxtal

# Define the reference template environments
xtal = pyxtal()
xtal.from_prototype('a-cristobalite')
print(xtal)

# Initialize the builder
bu = builder(['Si', "O"], [1, 2],
             db_file='sio2.db',
             log_file='sio2.log',
             verbose=True)
bu.set_descriptor_calculator(mykwargs={'rcut': 2.4})
bu.set_reference_enviroments(xtal.to_ase())
bu.set_criteria(CN={'Si': [4], 'O': [2]}, exclude_ii=True)
print(bu)

# Test pre-relaxatin on the known quartz structure
quartz = pyxtal()
quartz.from_prototype('a-quartz')
bu.optimize_xtal(quartz)

# Random Generation of using basin-hopping
bu.generate_xtal(spg=154,
                 wps=[['3a'], ['6c']],
                 niter=20,
                 early_quit=0.02)
\end{lstlisting}

The output should look like the following:
{\small
\begin{verbatim}
------MOF Builder------
System: Si1 O2 
Database: sio2.db
Log_file: sio2.log
Descriptor: SO3 descriptor with Cutoff: 2.400 
lmax: 4, nmax: 2, alpha: 1.500
Reference enviroments (2, 15)
Criterion_CN: {'Si': [4], 'O': [2]}
Criterion_cutoff: None
Criterion_exclude_ii: True

Test pre-relaxation on the existing quartz 
* 9 6 152 P3121  2.46 True 14.36 => 0.00 3a 6c 

Test random generation using basin-hopping
3 154 [['3a'], ['6c']] 0.2 5 0.02
* 9 6 154 P3221 2.46 0.000 3a 6c 
* 9 6 154 P3221 2.46 True   0.00 => 0.00 3a 6c 
\end{verbatim}
}

Clearly, it suggests our code can handle multiple environment in a straight manner for relaxation, as well as generative task by using the Basin Hopping approach. The structure generation process can also be extended to the use of AI generative models such as GAN and VAE. To enable this function in \texttt{LEGO-xtal} workflow, one can train a VAE or GAN model using a dataset of multi-component structures. The trained model can then be used to generate new candidate structures, which can subsequently undergo descriptor-based pre-relaxation and final relaxation \& energy ranking using GULP or MACE. This approach allows for the exploration of a wide range of multi-component crystal structures while ensuring that the generated structures are physically reasonable and energetically favorable.

\section{Impact of Training data}
\subsection{Pre-filtered Structures}
To construct the training dataset, we initially extracted a total of 150 sp$^2$ crystal structures from the SACADA database. Upon careful examination, we identified 10 structures that were unsuitable for inclusion due to one or more of the following criteria: (1) a large number of atoms per unit cell ($>$500), (2) an excessive number of Wyckoff sites ($>$8), or (3) a high number of irreducible variables ($>$24). These 10 structures were therefore excluded from the final training set. Details of the excluded structures are provided in Table~\ref{tab:pre_filtered_sp2}.

\begin{table*}[ht]
    \centering
    \caption{Summary of neglected sp$^2$ structures used for training.}
    \begin{tabular}{|c|c|c|c|c|c|}
        \hline
        \textbf{ID} & \textbf{\# Atoms} & \textbf{DOF} & \textbf{Space Group} & \textbf{List of Wyckoff Sites} & \textbf{VASP energy} \\
        & & & & & eV/atom\\
        \hline
        1 & 96 & 18 & 66 ($Cccm$) & 16m 16m 16m 8j 8j 8j 8k 8k 8k & -9.123 \\
        2 & 40 & 29 & 10 ($P$2/$m$) & 4o 4o 4o 4o 4o 2j 2j 2j 2j 2j 2k 2k 2k 2l 2l & -9.106 \\
        3 & 72 & 38 & 150 ($P$321) & 6g 6g 6g 6g 6g 6g 6g 6g 6g 6g 6g 6g & -9.051 \\
        4 & 24 & 19 & 10 ($P$2/$m$) & 4o 4o 4o 2l 2l 2l 2j 2j 2i & -9.047 \\
        5 & 512 & 11 & 227 ($Fd$-3$m$) & 32e 96g 96g 192i 96g & -8.852 \\
        6 & 8 & 30 & 1 ($P$1) & 1a 1a 1a 1a 1a 1a 1a 1a & -8.646 \\
        7 & 8 & 30 & 1 ($P$1) & 1a 1a 1a 1a 1a 1a 1a 1a & -8.353 \\
        8 & 672 & 22 & 203 ($Fd$-$3$) & 96g 96g 96g 96g 96g 96g 96g & -8.353 \\
        9 & 60 & 29 & 12 ($C$2/$m$) & 8j 4i 4i 8j 8j 4i 4i 8j 4i 8j & -8.415 \\
        10 & 64 & 26 & 24 ($I2_12_12_1$) & 8d 8d 8d 8d 8d 4c 4c 8d 8d & -8.086 \\
        \hline
    \end{tabular}
    \label{tab:pre_filtered_sp2}
\end{table*}

For reference, the lowest-energy sp$^2$ structure in the database—graphite—has a VASP energy of $-9.355$~eV/atom. Of the neglected structures, six exhibit relatively high energies ($>0.5$~eV/atom above graphite), making them less relevant for the generation of low-energy sp$^2$ allotropes. The remaining four structures, while lower in energy, are highly complex and were excluded for simplicity. In future work, we may plan to incorporate these complex structures into the training dataset to enhance the model's ability to generate a broader range of sp$^2$ allotropes.

\subsection{Data Augmentation}
The data augmentation comes from two sources:
\begin{enumerate}
    \item Subgroup representation is used to improve the number of unique symmetries
    \item Swap of coordinates to represent symmetry operation in each Wyckoff position 
\end{enumerate}

Conceptually, adding more unique symmetries are very important to help the model’s diversity. On the other hand, adding the swap of coordinates may be applied to the same (space group, Wyckoff positions) multiple times in current data augmentation strategy. So there may be a redundancy here. 

To verify our hypothesis, we prepared seven datasets (1) 3K with 139 unique symmetries; (2) 8K with 155 unique symmetries; (3) 3K with 320 unique symmetries; (4) 12K with 341 unique symmetries; 
(5) 60K with 419 unique symmetries; (6) 120K with 449 unique symmetries; (7) 240K with 485 unique symmetries to train the VAE-cont models. We then generated 100,000 samples from each model
and test their performances thereafter. The results are summarized in Table \ref{tab:augmentation}.

\begin{table}
    \centering
    \caption{Performance of VAE-cont models trained on datasets with different sizes and unique symmetries.}
    \begin{tabular}{|c|c|c|c|c|c}
        \hline
        \textbf{Train Size} & \textbf{Unique Symmetries} & $N_\text{Valid}$ & $N_\text{unique}$ & $N_\text{train}$ \\
        \hline
        3K   & 139 & 19637 & 1852 & 51 \\
        8K   & 155 & 18466 & 2105 & 50 \\
        3K   & 320 & 15958 & 2899 & 61 \\
        12K  & 341 & 18149 & 4059 & 73 \\
        60K  & 419 & 16524 & 4862 & 100 \\
        120K & 449 & 15330 & 5123 & 94 \\
        240K & 485 & 15052 & 5156 & 101 \\
        \hline
    \end{tabular}
    \label{tab:augmentation}
\end{table}

The results reveal a clear trend: as the training dataset size and the number of unique symmetries increase, the success rate for generating valid sp$^2$ structures decreases, likely due to increased data complexity. Conversely, models trained on smaller datasets tend to produce more duplicate structures, resulting in fewer valid unique sp$^2$ crystals and a higher rate of reproducing training data. Thus, a sufficiently large and diverse training dataset is essential for enhancing model performance in terms of both novelty and reproducibility.

Based on the test performances, we recommend using the 240k dataset. Although this dataset may include redundant information due to repeated permutations of the same Wyckoff positions, subgroup augmentation clearly enhances model performance. This is evident by directly comparing the results trained on 3K-139 and 3K-320 datasets, in which the latter (2899) generates more unique sp2 crystals than the former (1852). This is not surprising since the models have learned more symmetry examples from the training data. This approach enables the model to learn both subgroup representations and the statistical data distribution of Wyckoff sites in the given space group, thereby increasing the likelihood of generating crystals with subgroup symmetries.

Finally, subgroup augmentation is expected to be even more advantageous for multicomponent systems, as it facilitates learning structural analogues (e.g., from diamond to cubic boron nitride as described in our introduction section), thereby broadening the scope of generative crystal design.

\subsection{Subgroup Augmentation Details}
To perform subgroup augmentation, we utilized the \texttt{PyXtal} package \cite{pyxtal} to generate subgroups for each structure in our initial dataset using a generic way as described in Listing \ref{lst:subgroup}.

\begin{lstlisting}[language=Python, caption=A Python script to generate the subgroup structures, label=lst:subgroup]
# pip install pyxtal
from pyxtal import pyxtal

# load a graphite crystal
xtal=pyxtal()
xtal.from_prototype('graphite')

print("Derive subgroup graphite structures")
sub_t = xtal.subgroup(eps=0.01, group_type='t')
print("t_subgroup xtals", len(sub_t))

sub_k = xtal.subgroup(eps=0.01, group_type='k', 
                      max_cell=9)
print("k_subgroup xtals", len(sub_k))
\end{lstlisting}

In this work, we report the results by using a very small random distortion (0.001~\AA). Conceptually, the benefit of subgroup augmentation is mainly to encourage the model to try new choices for the discrete combinations of space group number and Wyckoff site choices. Hence, the numerical noises on continuous variables should not impact too significantly. To verity this hypothesis, we tried to apply a larger distortion up to (0.2~\AA) and train the generative models. There seems to be no significant impact on the results, thus confirming this hypothesis.

\section{Analysis of Generative Model Performance}
\subsection{Comparison with Random Sampling}
Our data augmentation strategy leverages subgroup representation to enhance the diversity of generated structures, particularly by increasing the number of unique symmetries. However, it is important to assess whether a generative model trained on lattice parameters and reduced Wyckoff coordinates truly outperforms random sampling. 

To address this concern, we performed a sanity check using 1,000 samples. When lattice parameters and Wyckoff positions were randomly generated for a fixed space group and Wyckoff site, only 8 unique valid structures were obtained. In contrast, our trained VAE model produced 108 unique structures under the same conditions. Furthermore, fully randomizing the tabular representation (without respecting space group or Wyckoff constraints) yielded only 2 valid structures.

These results confirm that learning a distribution over lattice and reduced coordinates substantially improves both structural diversity and validity compared to naive random sampling. Nevertheless, we acknowledge that the model may be overfitted and could generate data within a limited range. Therefore, we retain the option to allow random sampling for further exploration.

\subsection{Saturation Analysis}
\begin{figure}
    \centering
    \includegraphics[width=0.5\textwidth]{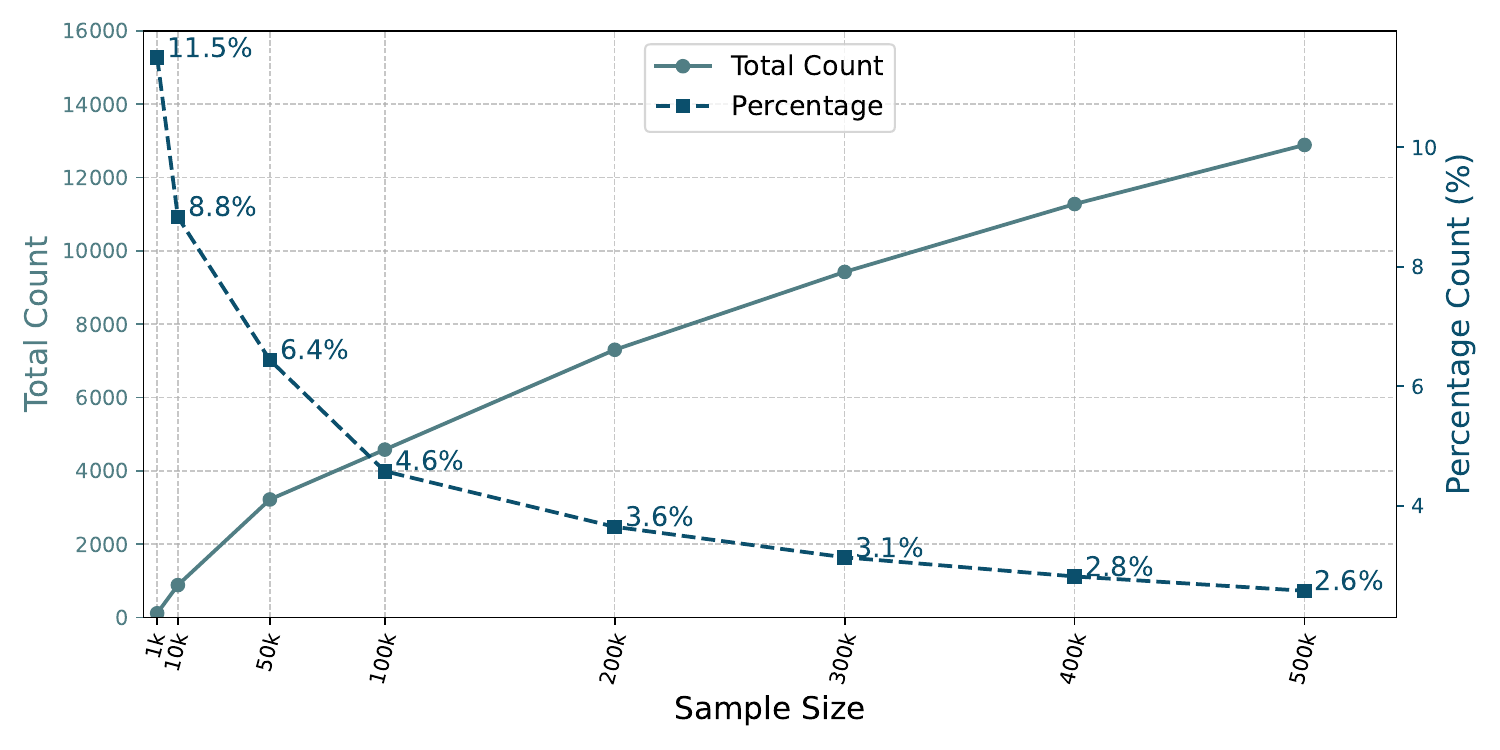}
    \caption{\textbf{Saturation Analysis.} The number of unique sp$^2$ structures generated by the VAE model as a function of the total number of samples. The curve shows that while the rate of discovering new unique structures decreases with more samples, saturation is not reached, indicating continued diversity in generated structures.}   \label{fig:Saturation}
\end{figure}

As summarized in Table \ref{tab_summary} of the main text, approximately 4,000 - 6,000 unique sp$^2$ structures per 100,000 samples appears to be the upper limit observed across all strategies tested. However, true diversity does not fully saturate within this range. By generating increasingly larger sample sets using the VAE model trained on the v1-60k-cont dataset (specifically, batches of 1k, 10k, 50k, 100k, 200k, 300k, 400k, and 500k samples), we observe a continued increase in the total number of unique structures. Figure~\ref{fig:Saturation} illustrates this trend: although the rate of discovering new unique sp$^2$ structures decreases as more samples are generated, saturation is not reached, indicating persistent diversity in the generated dataset.

\section{An Extended List of sp\texorpdfstring{$^2$}{2} Allotropes}
The full list of structural information can be found at \url{https://lego-crystal.onrender.com}. Below we highlight several representative structures and their dynamical stability.

\subsection{The 2D sp\textsuperscript{2} Allotropes}
\begin{figure*}[htbp]
    \centering
    \includegraphics[width=0.8\textwidth]{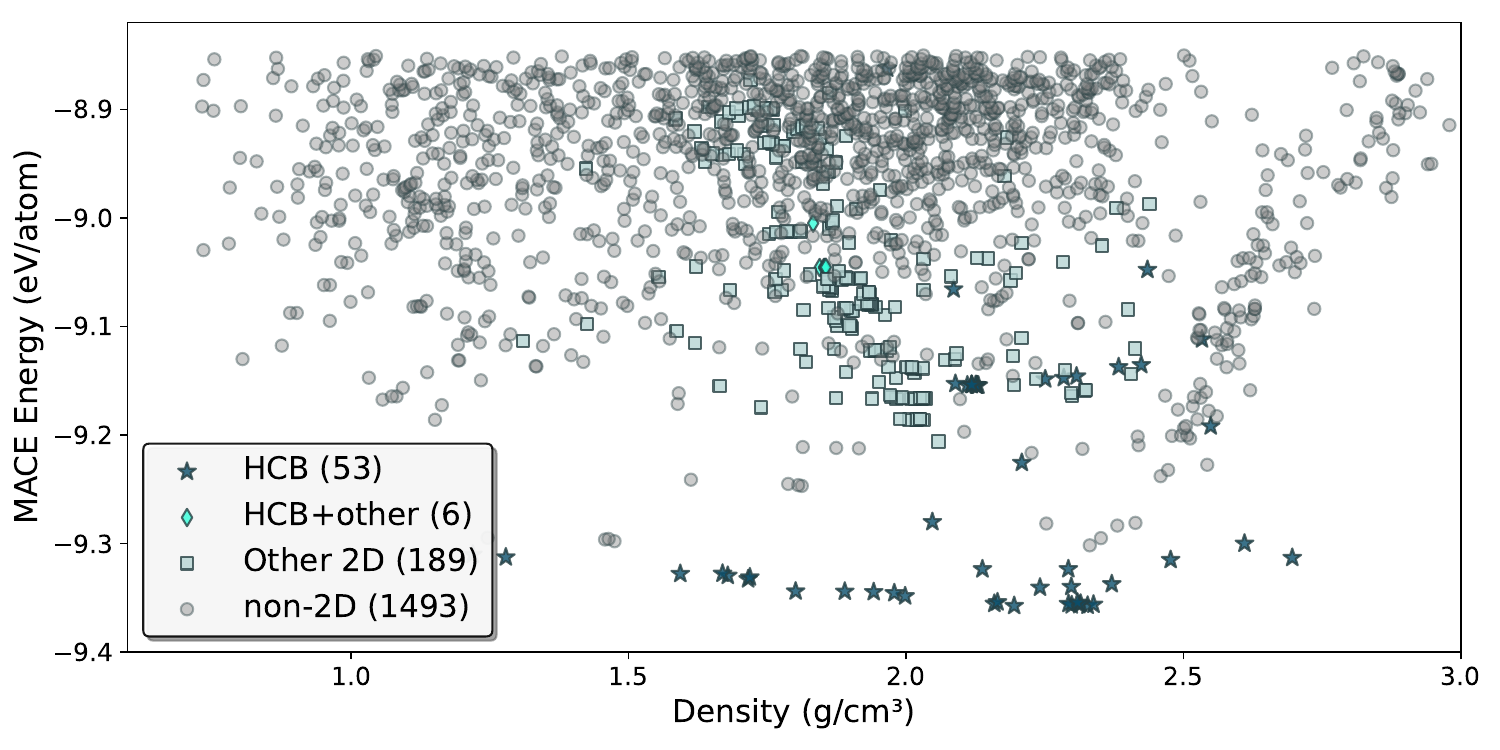}
    \caption{\textbf{Energy versus density plot for 2D sp$^2$ allotropes.} The distribution of energies and densities for 248 unique 2D sp$^2$ carbon structures generated by the VAE model. The plot highlights the diversity of topologies, including graphene stacking variants and other low-energy motifs, as compared to the ground state graphite layer.}
    \label{fig:2d-sp2}
\end{figure*}

Out of 1741 structures from our database, there are a total of 248 2D sp2 allotropes, including (1) 53 structures featured by different kinds of graphene (hcb topology) stacking variants; (2) 6 hybrid stacking between hcb and other 2D layer motifs; (3) 189 other 2D motifs. Among the 189 other 2D motifs, they include many well known topologies (e.g., hnb, hnc, hae, hae, etc.), as well as many defective-graphene structures with various types of non-hexagonal rings (e.g., 5-7, 5-8, 4-6, 4-8, etc.). The energy versus MACE energy plot is shown in Figure \ref{fig:2d-sp2}.  

\begin{figure*}[htbp]
    \centering
    \includegraphics[width=0.99\textwidth]{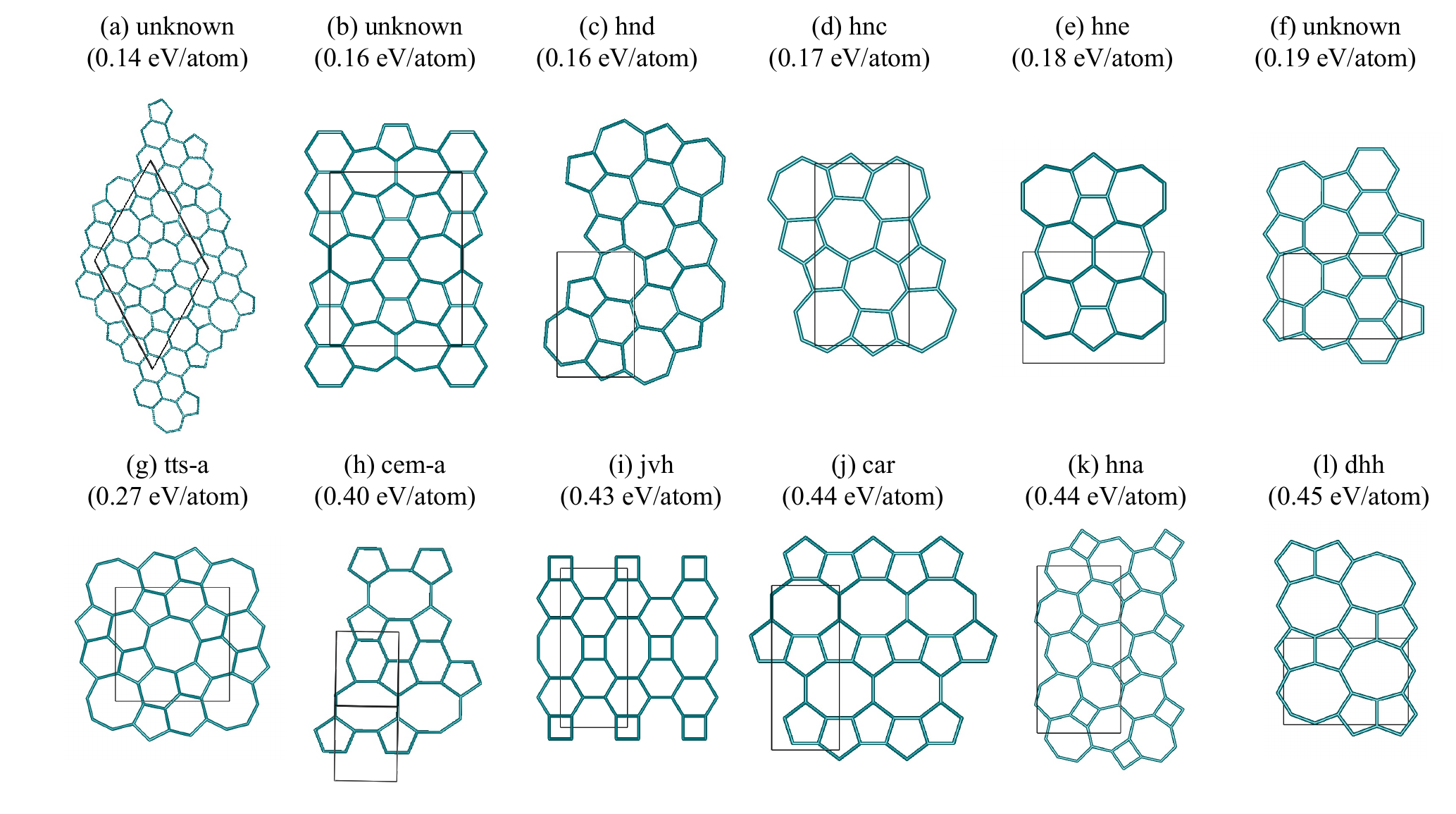}
    \caption{\textbf{An extended list of of low-energy 2D carbon sp$^2$ layers.} The known topologies are denoted according to notation used in RCSR \cite{rcsr}. The energy is provided based on the MACE model as compared to the ground state graphite (hcb) layer.}
    \label{fig:extended-2d}
\end{figure*}

Fig. \ref{fig:extended-2d} displays various 2D layer motifs found in the LEGO-sp$^2$ database. While these motifs possess higher energies than the ground state graphite layer, they represent distinct topological arrangements that could exhibit interesting electronic properties worthy of further investigation.

\subsection{The Low-energy 3D and 0D sp\textsuperscript{2} Allotropes}

\begin{figure*}[htbp]
    \centering
    \includegraphics[width=0.85\textwidth]{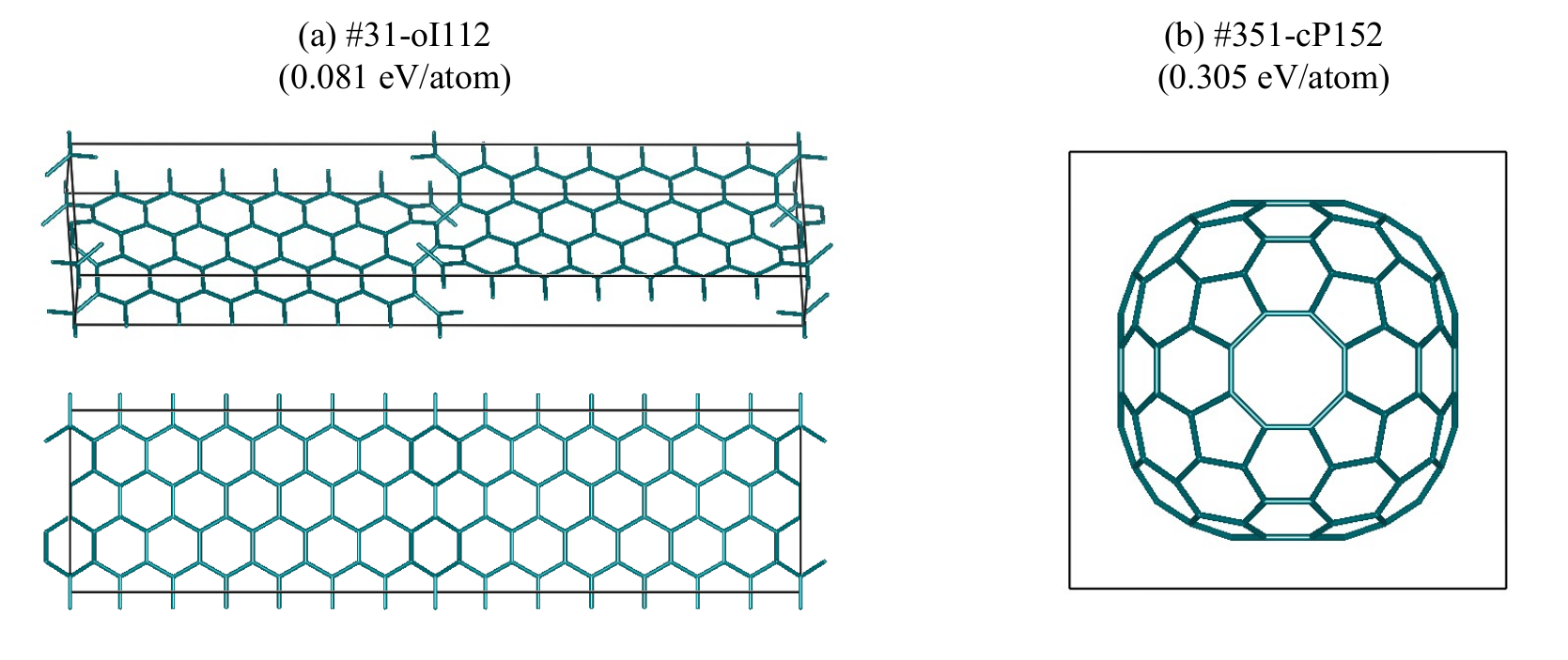}
    \caption{\textbf{Two low-energy structures.} (a) the lowest-energy 3D structure with alterative graphite layer stacking; (b) a 0D structure with the C$_{152}$ cage. The indices in the labels correspond to the entry ids in \url{https://lego-crystal.onrender.com} and the energy values are shown as the relative DFT-r$^2$SCAN energies as compared to the ground state graphite structure.}
    \label{fig:extended-lowE}
\end{figure*}

Fig. \ref{fig:extended-lowE} displays two notable structures discussed in the main text. Fig. \ref{fig:extended-lowE}a shows the lowest-energy 3D sp$^2$ allotrope, featuring alternating graphite layers connected by single bonds. These structures exhibit distinctive electronic properties due to their unique connectivity pattern \cite{zhao2020family}. Fig. \ref{fig:extended-lowE}b shows a 0D cage structure containing 152 carbon atoms, which represents a novel addition to the fullerene family. 

\subsection{Phonon Calculations for Selected Structures}
\begin{figure*}[htbp]
    \centering
    \includegraphics[width=0.85\textwidth]{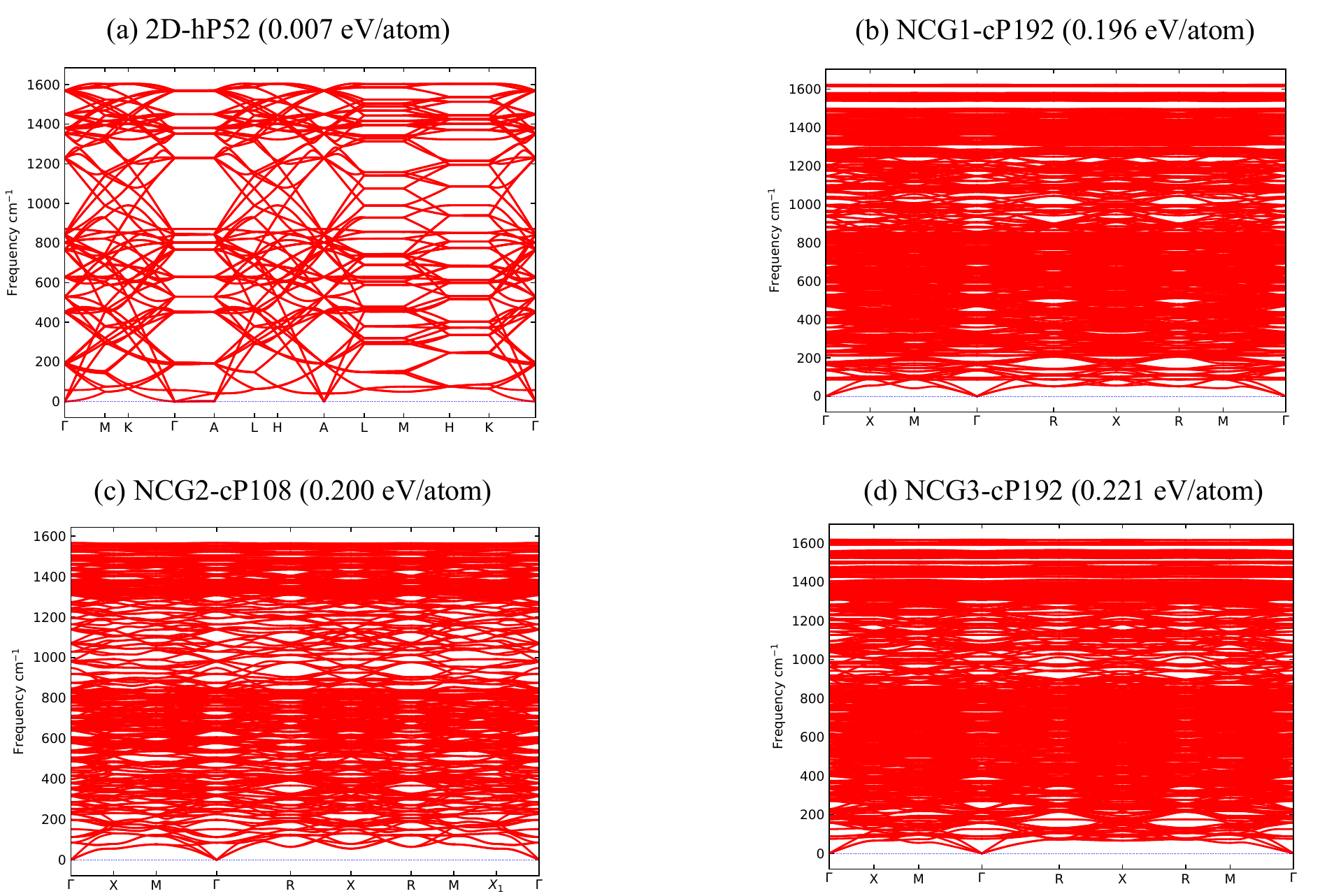}
    \caption{\textbf{Calculated phonon dispersions for a few selected low-energy structures at the PBE level of theory.} (a) is the twisted graphene shown in Figure 6b. (b)-(d) are NCG1-3 structures shown in Figure 7.}
    \label{fig:phonon}
\end{figure*}

Fig. \ref{fig:phonon} displays the calculated phonon for 4 representative low-energy sp$^2$ allotropes that are reported in this work for the first time, including (a) 2D-hP52, (b) NCG1-cP192, (c) NCG2-cP108 and (d) NCG3-cP192.

\section*{REFERENCES}
\bibliography{ref}

\end{document}